\pdfoutput=1
\documentclass{article} 
\usepackage[final]{colm2025_conference} 

\usepackage{microtype}
\usepackage{hyperref}
\usepackage{url}
\usepackage{booktabs}
\usepackage{wrapfig}

\usepackage{lineno}

\definecolor{darkblue}{rgb}{0, 0, 0.5}
\hypersetup{colorlinks=true, citecolor=darkblue, linkcolor=darkblue, urlcolor=darkblue}

\usepackage{xspace}
\usepackage{amsmath}
\usepackage{lmodern}
\usepackage[normalem]{ulem}
\usepackage{tcolorbox}
\usepackage{subcaption}
\usepackage{multirow}
\usepackage{paralist}
\usepackage{xparse}
\usepackage{listings}

\newcommand{\gpt}{\textsf{GPT}\xspace}
\newcommand{\codellama}{\textsf{Code Llama}\xspace}
\newcommand{\llama}{\textsf{Llama}\xspace}
\newcommand{\verieql}{\textsf{VeriEQL}\xspace}
\newcommand{\spider}{\textsf{Spider}\xspace}
\newcommand{\spiderdin}{\textsf{Spider+DIN}\xspace}
\newcommand{\calcite}{\textsf{Calcite}\xspace}
\newcommand{\exams}{\textsf{SQLEquiQuest}\xspace}
\newcommand{\geminipro}{\textsf{Gemini-Pro}\xspace}

\newcommand{\sqlsolver}{\textsf{SQLSolver}\xspace}

\newcommand{\raj}[1]{#1}

\newcommand{\eat}[1]{}
\definecolor{magnolia}{rgb}{0.97, 0.96, 1.0}
\definecolor{verdigris}{rgb}{0.26, 0.7, 0.68}
\definecolor{codegray}{rgb}{0.5, 0.5, 0.5}
\definecolor{codepurple}{rgb}{0.58, 0, 0.82}

\lstdefinestyle{Sqlstyle}{
    language=SQL,
    backgroundcolor=\color{magnolia},
    commentstyle=\color{verdigris},
    keywordstyle=\color{magenta},
    numberstyle=\tiny\color{black},
    stringstyle=\color{codepurple},
    basicstyle=\ttfamily\tiny,
    breakatwhitespace=false,
    breaklines=true,
    captionpos=b,
    keepspaces=false,
    showspaces=false,
    showstringspaces=false,
    showtabs=false,
    tabsize=2
}

\title{Can the Rookies Cut the Tough Cookie? Exploring the Use of
LLMs for SQL Equivalence Checking}

\author{Rajat Singh \\
    Dept. of Computer Science and Engineering\\
    Indian Institute of Technology\\
    Delhi, 110016, India \\
    \texttt{rajat.singh@cse.iitd.ac.in}\\
    \And
    Srikanta Bedathur \\
    Dept. of Computer Science and Engineering\\
    Indian Institute of Technology\\
    Delhi, 110016, India \\
    \texttt{srikanta@cse.iitd.ac.in}\\
}

\begin{document}

\ifcolmsubmission
\linenumbers
\fi

\maketitle

\begin{abstract}
Equivalence checking of SQL queries is an intractable problem often encountered in settings ranging from grading SQL submissions to debugging query optimizers. Despite recent work toward developing practical solutions, only simple queries written using a small subset of SQL are supported, leaving the equivalence checking of sophisticated SQL queries at the mercy of intensive, potentially error-prone, manual analysis. In this paper, we explore how LLMs can be used to reason with SQL queries to address this challenging problem. Towards this, we introduce a novel, realistic, and sufficiently complex benchmark called \exams for SQL query equivalence checking that reflects real-world settings. We establish strong baselines for SQL equivalence checking by leveraging the ability of LLMs to reason with SQL queries. We conduct a detailed evaluation of several state-of-the-art LLMs using various prompting strategies and carefully constructed in-context learning examples, including logical plans generated by SQL query processors. Our empirical evaluation shows that LLMs go well beyond the current capabilities of formal models for SQL equivalence, going from a mere 30\% supported query pairs to full coverage, achieving up to 82\% accuracy on \spiderdin. However, a critical limitation of LLMs revealed by our analysis is that they exhibit a strong bias for equivalence predictions, with consistently poor performance over non-equivalent pairs, opening a new direction for potential future research. 

\end{abstract}

\renewcommand{\thesubfigure}{\Roman{subfigure}}

\section{Introduction}
\label{sec:intro}
Determining if two arbitrary SQL queries are equivalent -- i.e., they generate the same output when executed on any database instance of the given schema -- is a classical ``tough cookie'' to cut with well-established undecidability results~\citep{ChandraMerlin1977,Abiteboul1995FoundationsOD,Ioannidis1995ContainmentOC}. However, the need to check equivalence appears in many practical scenarios, such as evaluating SQL assignment submissions in a university classroom setting~\citep{Chandra2014DataGF}, verifying text-to-SQL results, and ensuring correct query rewriting rules in a query optimizer~\citep{10.1145/356924.356928}. Current ad hoc approaches, such as executing queries on an (often small) database instance and comparing the results~\citep{He2024VeriEQLBE,Yin2020TaBERTPF}, or measuring the similarity between two SQL statements using an edit-distance-based method~\citep{Yin2020TaBERTPF}, tend to be highly error-prone. 

Consequently, significant effort has been devoted to exploring computable query equivalence for specific sub-classes of relational queries and under certain assumptions ~\citep{ChandraMerlin1977,Zhou2022SPESAS,Ding2023ProvingQE,Chaudhuri1993OptimizationOR,Wang2023TowardsPS,Zhou2019AutomatedVO,Haynes2023GEqOMS,Wang2022WeTuneAD,Chu2017CosetteAA,He2024VeriEQLBE,Chu2016HoTTSQLPQ,Cohen2006EquivalenceOQ}. Despite the long list of technically sophisticated results \& well-crafted tools, we are still far from building a model that can be reliably used in production to verify the equivalence of SQL queries of arbitrary complexity. For example, consider the two SQL queries in Figure~\ref{fig:example_SQL_toughcookie}. These queries are part of an assignment in a university graduate-level course, where Query\_1 and Query\_2 are non-equivalent. Their non-equivalence is not immediately obvious, even to those familiar with SQL. Moreover, SQL equivalence-checking tools such as \verieql~\citep{He2024VeriEQLBE}, \sqlsolver~\citep{Ding2023ProvingQE}, and \textsf{Cosette}~\citep{Chu2017CosetteAA} either do not support this pair of queries or fail to provide an explanation for why the queries are equivalent or non-equivalent. In fact, in the novel \exams benchmark we present in this paper (cf. Section~\ref{sec:Exams_iitd}), only about $30\%$ pairs are supported by \sqlsolver~\citep{Zhao2023LLMSQLSolverCL} and mere $2.8\%$ by \verieql~\citep{He2024VeriEQLBE} -- two recent SQL equivalence checkers (for details, see Section~\ref{sec:results_and_analysis}). 

\begin{figure*}[t]
    \centering
    \begin{subfigure}[c]{0.52\textwidth}
        \centering
        \begin{minipage}{\columnwidth}
        \begin{lstlisting}[style=Sqlstyle]
 WITH result AS
 (
     SELECT
         playerid,
         SUM (COALESCE (cs ,0)) AS total_caught_stealing
     FROM batting
     GROUP BY batting.playerid
     ORDER BY total_caught_stealing DESC 
 )
 SELECT 
     result.playerid,
     COALESCE (people.namefirst, '') AS firstname,
     COALESCE (people.namelast, '') AS lastname,
     result.total_caught_stealing
 FROM result
 JOIN people
 ON result.playerid = people.playerid
 ORDER BY 
     total_caught_stealing DESC,
     namefirst ASC,
     namelast ASC,
     playerid ASC LIMIT 10;

        \end{lstlisting}
        \end{minipage}
        \caption{Query\_1}
        \label{fig:image1_q1}
    \end{subfigure}
    \hfill
    \begin{subfigure}[c]{0.47\textwidth}
        \centering
        \begin{minipage}{\columnwidth}
        \begin{lstlisting}[style=Sqlstyle]









        
 SELECT 
     people.playerid, 
     namefirst AS firstname, 
     namelast AS lastname, 
     cs AS total_caught_stealing 
 FROM people 
 JOIN batting ON people.playerid = batting.playerid 
 ORDER BY 
     COALESCE(cs, 0) DESC, 
     COALESCE(namefirst, '') ASC, 
     COALESCE(namelast, '') ASC, 
     people.playerid ASC LIMIT 10; 


        \end{lstlisting}
        \end{minipage}
        \caption{Query\_2}
        \label{fig:image2_q2}
    \end{subfigure}
    \caption{Example of a non-equivalent pair of SQL queries from a database assignment question (\exams dataset).}
\label{fig:example_SQL_toughcookie}
\end{figure*}

\noindent \textbf{Large Language Models for SQL: }
Large language models (LLMs), among other capabilities, have demonstrated exceptional performance in tasks such as text-to-SQL~\citep{Liu2023ACE,Pourreza2023DINSQLDI,Gao2023TexttoSQLEB,Xie2022UnifiedSKGUA}, and also in generating natural language explanations of code segments -- including SQL -- with remarkable accuracy~\citep{Sun2023APL,Bhattacharya2023ExploringLL,Geng2023LargeLM,Gao2023WhatMG}. By leveraging these capabilities of LLMs, we aim to develop a robust solution for determining the equivalence of SQL queries, potentially overcoming the limitations of existing formal methods and tools.

In this work, we will treat LLMs as black-box experts -- advanced artificial intelligence (AI) models -- capable of understanding and generating human-like text. Given the extensive corpus used during their training, it is highly likely that these LLMs have been exposed to datasets used in existing work. Furthermore, prevailing benchmarks, such as SPIDER+DIN, primarily consist of relatively simple SQL queries that do not accurately represent the complexity and diversity of real-world SQL queries encountered in practical applications. This underscores the need for novel, diverse, and complex datasets to effectively evaluate LLMs on the task of equivalence of SQL queries.

\subsection{Contributions}
In summary, this paper makes the following key contributions:
\begin{itemize}

    \item Presents a practical and challenging SQL equivalence benchmark called \exams, derived from advanced SQL assignments at a graduate-level DBMS course. We believe this novel benchmark will further spur research activity in this task. 

    \item Introduces the task of SQL query equivalence to evaluate the SQL reasoning abilities of LLMs, and establishes a strong baseline performance across a number of LLMs, using SQL logical plans in context. The experimental results achieved up to 70\% coverage gains over traditional SQL equivalence checkers,  with nearly 82\% accuracy on \spiderdin dataset and 61\% accuracy on \exams dataset.
    
    
\end{itemize}

\noindent
\textbf{Organization}: The structure of this paper is as follows: Section~\ref{sec:background} reviews related work, focusing on traditional approaches to SQL equivalence checking and recent advancements leveraging LLMs. Section~\ref{sec:dataset} discusses the lack of benchmarks for SQL equivalence checking, and presents our \exams benchmark in addition to a recently proposed approach for synthetically generated \spiderdin dataset based on well-known \spider benchmark for SQL generation. Section~\ref{sec:proposed_model} introduces various prompting strategies for addressing SQL query equivalence tasks. Section~\ref{sec:results_and_analysis} presents the results and analyzes the performance and effectiveness of our proposed approach. Lastly, Section~\ref{sec:conclusion} concludes the paper with a summary of the findings and a discussion of promising directions for future research.

\noindent
\textbf{Code}: The code and instructions for running it can be found in the following link: \hyperlink{https://github.com/rajatb115/LLMs-for-SQL-Equivalence-Checking.git}{https://github.com/rajatb115/LLMs-for-SQL-Equivalence-Checking.git}

\section{Related Work}
\label{sec:background}
Determining the equivalence of two arbitrary SQL queries is an undecidable problem \citep{Mostowski1950ReviewBA,Abiteboul1995FoundationsOD,Ioannidis1995ContainmentOC}. Nevertheless, significant research has focused on SQL query equivalence under specific assumptions~\citep{Zhou2022SPESAS,Ding2023ProvingQE,Chaudhuri1993OptimizationOR,Wang2023TowardsPS,Zhou2019AutomatedVO,Haynes2023GEqOMS,Wang2022WeTuneAD,Chu2017CosetteAA}. This particular problem has been extensively explored under two classes, i.e., set semantics~\citep{Zhou2019AutomatedVO,ChandraMerlin1977,Chekuri1997ConjunctiveQC,Johnson1983OptimizingCQ}, where tuples are duplicate-free, and bag semantics~\citep{UDP,Wang2022WeTuneAD,Ding2023ProvingQE,Zhou2022SPESAS,Chaudhuri1993OptimizationOR}, which allows duplicate tuples. 


Traditionally, researchers have used two distinct approaches to investigate the problem of query equivalence: the algebraic approach~\citep{UDP,Chu2017CosetteAA,Wang2022WeTuneAD,Ding2023ProvingQE} and the symbolic approach~\citep{Zhou2019AutomatedVO,Zhou2022SPESAS}. Each method has its own set of advantages and disadvantages. Notable works such as COSETTE~\citep{Chu2017CosetteAA}, UDP~\citep{UDP}, WeTune~\citep{Wang2022WeTuneAD}, and \sqlsolver~\citep{Ding2023ProvingQE} employ the algebraic approach for the SQL query equivalence task. The algebraic approach involves converting SQL queries into algebraic expressions using an abstract syntax tree (AST). These expressions are then normalized, and their equivalence is proven using algebraic theory. However, this methodology has limitations, particularly in demonstrating equivalence for SQL queries with complex arithmetic expressions, three-valued logic, and specific combinations of SQL operators, such as certain aggregate functions in conjunction with OUTER JOIN.

In contrast, the symbolic approach, as seen in works like EQUITAS~\citep{Zhou2019AutomatedVO} and SPES~\citep{Zhou2022SPESAS}, involves transforming SQL queries into \textit{first-order logic} (FOL) using a collection of symbolic variables. Satisfiability Modulo Theories (SMT) is then used to verify the equivalence of these FOL formulas. The symbolic approach enables more powerful semantic reasoning than the algebraic approach and addresses many limitations. However, symbolic approaches are generally limited to proving SQL equivalence under set semantics, whereas real-world SQL queries often follow bag semantics.

Closely related to our work is that of LLM-SQL-Solver~\citep{Zhao2023LLMSQLSolverCL}, which also explores the potential of LLMs in evaluating the equivalence of SQL queries by identifying the minimal changes required for a given SQL query to make it equivalent to a baseline query using LLMs. We use their novel benchmark design, where DIN-SQL~\citep{Pourreza2023DINSQLDI}, a powerful text-to-SQL model, was used in the Spider benchmark to generate multiple correct SQL queries for the given natural language query description (see Appendix, Section~\ref{app:dataset}). In our work, we aim to address several questions that remain unanswered by LLM-SQL-Solver. Specifically, we investigate the effects of fine-tuning LLMs to determine SQL query equivalence, as well as a broader range of prompting techniques with an unoptimized SQL Logical Plan (LP) to improve the performance of LLMs. We also discovered that the LLM-SQL-Solver evaluation metric is potentially misleading, which we correct and report updated performance numbers (see Section~\ref{performance_of_llm}).

\noindent\textbf{SQL Query Equivalence:}
For the context of this paper, we define the notion of SQL equivalence as follows: Let \( Q_1 \) and \( Q_2 \) be two SQL queries, and let \( D \) be a given database schema. The queries \( Q_1 \) and \( Q_2 \) are considered ``\textit{Equivalent}" if and only if, for every possible instance of the schema \( D \), executing both queries produces identical results. That is, for all valid database instances \( I \) conforming to \( D \), the outputs of \( Q_1(I) \) and \( Q_2(I) \) must be identical. Conversely, \( Q_1 \) and \( Q_2 \) are considered ``\textit{Non-Equivalent}" if there exists at least one valid database instance \( I \) conforming to \( D \) for which \( Q_1(I) \) and \( Q_2(I) \) produce different results.

\section{Benchmarks for SQL Equivalence Checking}
\label{sec:dataset}
\begin{figure}[t]
    \centering
    \begin{subfigure}[b]{0.29\textwidth}
        \centering
        \includegraphics[width=\textwidth]{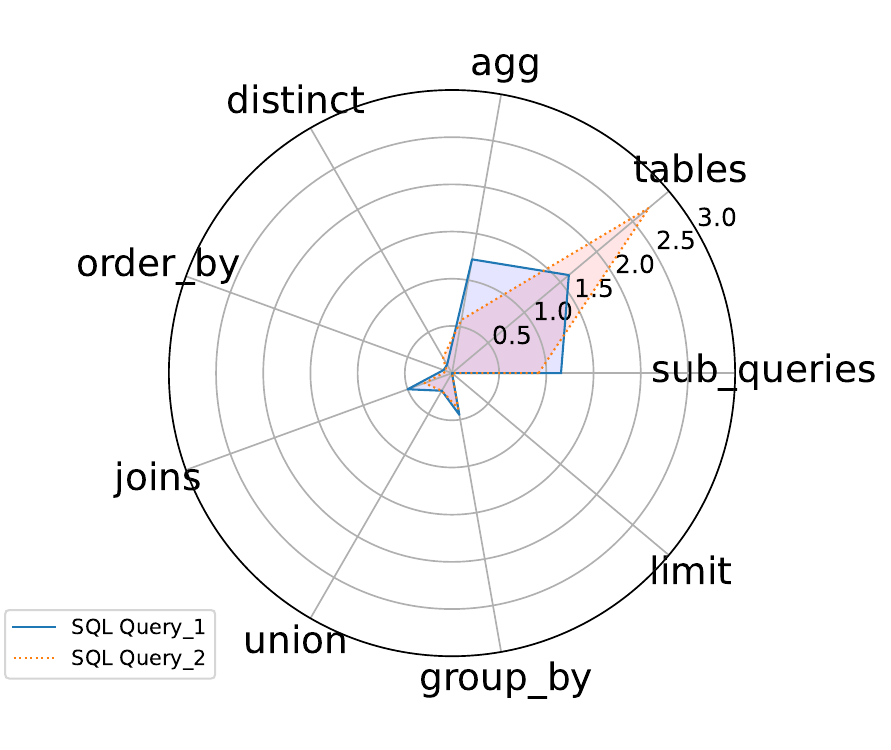}
        \caption{\calcite}
        \label{fig:calcite_complexity}
    \end{subfigure}
    \hfill
    \begin{subfigure}[b]{0.32\textwidth}
        \centering
        \includegraphics[width=\textwidth]{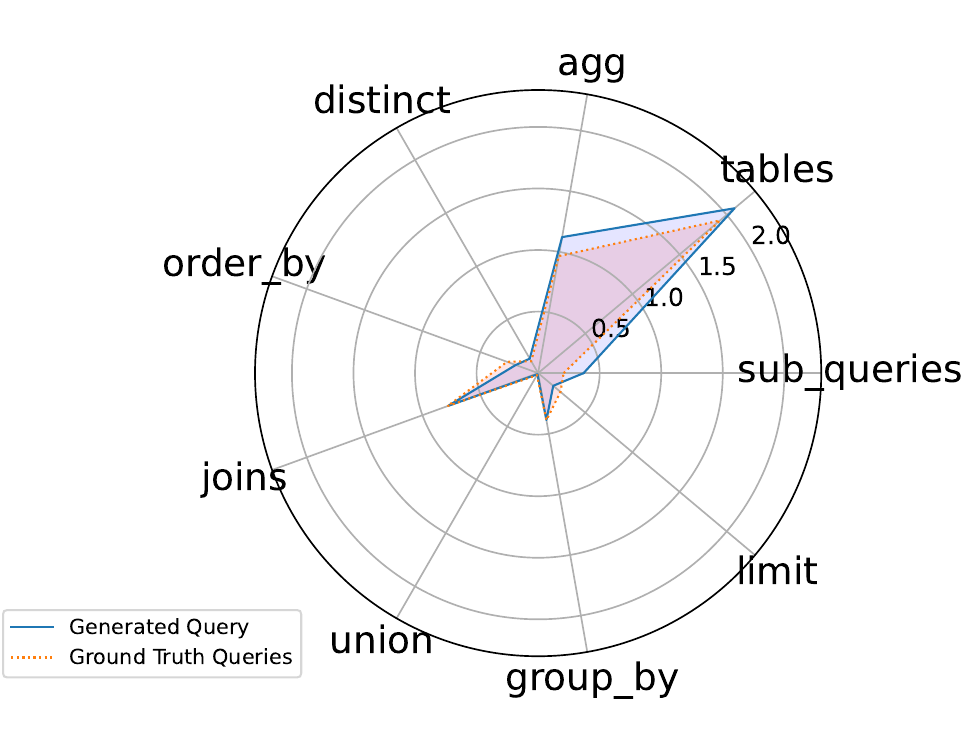}
        \caption{\spiderdin}
        \label{fig:spider_din_complexity}
    \end{subfigure}
    \hfill
    \begin{subfigure}[b]{0.355\textwidth}
        \centering
        \includegraphics[width=\textwidth]{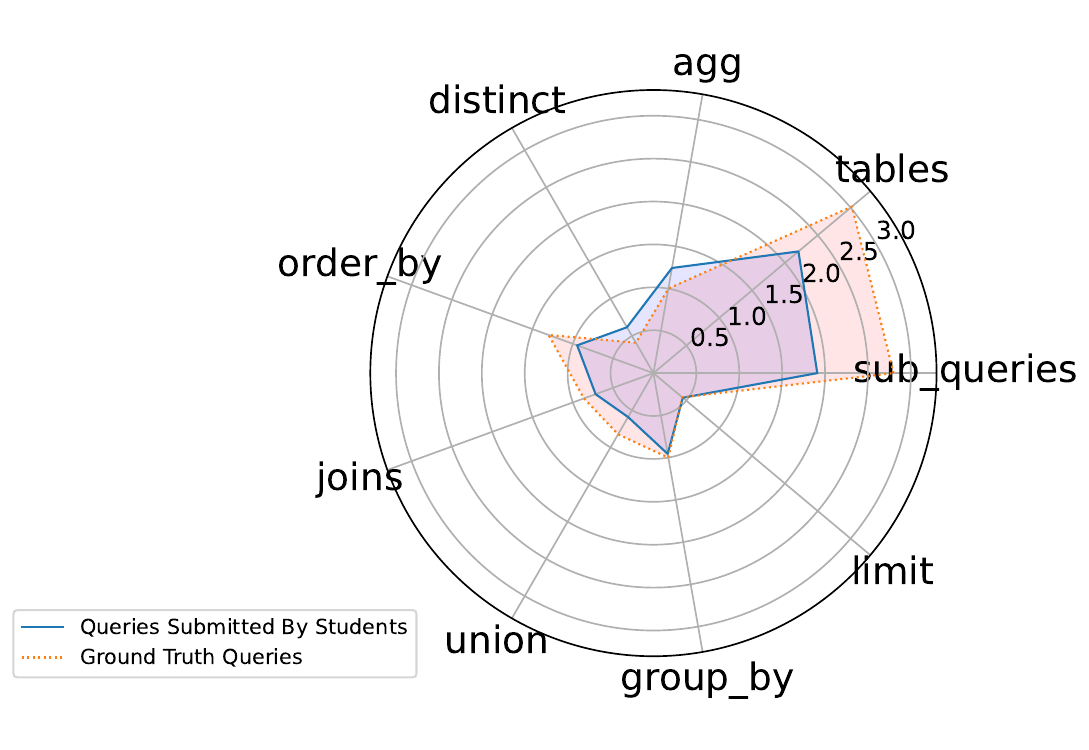}
        \caption{\exams}
        \label{fig:sqlequiquest_complexity}
    \end{subfigure}
    \caption{Distribution of various SQL features in \textit{\calcite}, \textit{\spiderdin}, and \textit{\exams} dataset. The radial axis represents the average number of corresponding feature in a single query in the benchmark. Also note that \textit{\calcite} contains only equivalent pairs of SQL queries.}
    \label{fig:all_spider_sqlequiquest}
\end{figure}

Since SQL query equivalence checking is a fairly important problem for database developers, a considerable amount of research has been undertaken to develop various ways of addressing the issue. Despite this, there is a dearth of realistic benchmarks. There are smaller, narrowly focused datasets such as the ``Exams" benchmark~\citep{Chu2017CosetteAA} or a larger collection of queries such as those of the \calcite~\citep{calcite} optimizer team, which are repurposed to verify that the formal equivalence checking algorithms are indeed behaving as expected. Note that the Calcite dataset only contains equivalent pairs of SQL (before and after operator rewriting by the Calcite optimizer). \calcite dataset consists of 232 query pair samples.

A recent work~\citep{Zhao2023LLMSQLSolverCL} proposed a novel approach for generating a large collection of SQL queries using LLMs on the Spider dataset~\citep{yu2018spider}. The Spider dataset contains a collection of paired natural-language questions and equivalent SQL queries --along with the schema information and small-scale database corresponding to the query. One can generate SQL queries for each natural-language question using text-to-SQL models~\citep{Pourreza2023DINSQLDI,Gao2023TexttoSQLEB}. In this work, we used \spiderdin, as proposed by~\citep{Zhao2023LLMSQLSolverCL}, which leverage~\citep{Pourreza2023DINSQLDI} to generate additional SQL queries for each sample. Equivalence between the ground truth and the generated queries was assessed using test-suit-sql-eval~\citep{test-suite-sql-eval}. The generated dataset consists of 1034 SQL query pairs, of which 460 are exact matches. Among the remaining 574 query pairs, 385 are equivalent, while 189 are non-equivalent. 

Figures~\ref{fig:calcite_complexity} and~\ref{fig:spider_din_complexity} depict the complexity of the \calcite and \spiderdin datasets, respectively, measured by the average number of constraints per SQL query. The results indicate that both datasets are relatively simple, implying that they may not adequately capture the intricacies of real-world SQL queries encountered in practical applications. Furthermore, since \calcite and \spider are open-source datasets, it is very likely the they have already been exposed to LLMs during their training, potentially limiting their effectiveness in evaluating the true generalization capabilities of models. To address these limitations, we introduce \exams, a new benchmark designed to better reflect the complexity and diversity of real-world SQL queries, providing a more robust evaluation framework for LLM performance.
        
\subsection{\textbf{{\bfseries \exams} - a Benchmark of Undergraduate SQL Assignment Submissions}}
\label{sec:Exams_iitd}

Earlier work by~\citep{Chu2017CosetteAA} introduced the \textit{``Exams"} dataset, which comprises a handful of simple SQL query pairs curated from University of Washington semester exams. Building upon this foundation, we present \exams, a novel and extensive SQL query equivalence benchmark designed to capture greater complexity and diversity (Figure~\ref{fig:sqlequiquest_complexity}). The \exams dataset is constructed using five complex natural language (NL) questions provided to students as part of a database management systems (DBMS) take-home assignment conducted during the spring 2023-24 semester. The assignment required students to interpret NL descriptions specifying the required information, analyze a given database schema, and formulate appropriate SQL queries to satisfy the stated information needs.

\begin{wraptable}{r}{0.5\textwidth}
    \centering
    \resizebox{0.45\textwidth}{!}{
    \begin{tabular}{ccc}
    \toprule
        \textbf{Question Number} & \textbf{Equivalent} & \textbf{Non-Equivalent} \\
        \midrule
        Question 1 & 102 & 8 \\
        Question 2 & 102 & 7 \\
        Question 3 & 77 & 32 \\
        Question 4 & 16 & 86 \\
        Question 5 & 10 & 59 \\
        \bottomrule
    \end{tabular}
    }
    \caption{\exams dataset sample distribution.}
    \label{tab:exams_iitd_ground_truth}
\end{wraptable}

To ensure originality and mitigate potential exposure to LLMs, instructors carefully designed the schema and specifications, preventing their prior inclusion in the LLM training corpora. The ground-truth queries, written by instructors and verified by teaching assistants, exhibit real-world SQL complexities rarely seen in existing benchmarks. Student-submitted queries were systematically evaluated by teaching assistants using a large database instance designed to account for all potential corner cases. Two SQL queries were considered equivalent if they produced identical results when executed on the database instance; otherwise, they were classified as non-equivalent. The distribution of equivalent and non-equivalent submissions for each question in \spiderdin is presented in Table~\ref{tab:exams_iitd_ground_truth}.

The questions in the \spiderdin dataset span a range of difficulty levels, incorporating features such as sub-queries, joins, result ordering, and graph-based path queries. Question 1, the simplest in the set, received 102 correct submissions, whereas Question 5, the most challenging, had only 10 correct submissions. While Question 1 involves table joins and sub-queries with aggregate functions and ordering the result, Question 5 requires constructing a graph and identifying a path of a specified length, significantly increasing its complexity (Refer Appendix Section~\ref{app:ground_truth_queries} for further details). Thus, by incorporating queries with intricate constraints, nested structures, and diverse SQL constructs, \exams serve as a more representative benchmark for evaluating the SQL equivalence capabilities of LLMs.



\section{Prompting Strategies} 
\label{sec:proposed_model}

This section presents the prompting strategies used by LLMs for SQL equivalence checking tasks, categorized into two types: (i) \textbf{Single-Stage Prompting} and (ii) \textbf{Multi-Stage Prompting}. The exact prompts used in our experiments are present in the Appendix (for details, see Section~\ref{app:promting_strategies}).

\subsection{Single-Stage Prompting}
\label{single_stage_prompting}

The single-stage prompting strategy provides all the necessary information within a single prompt, guiding the LLM toward the desired output without requiring further interaction. Depending on the type or amount of information provided in the prompt, this approach can be categorized into different prompting techniques (Figure~\ref{P:method_1_2_4}):
\begin{inparadesc}
    \item[P1. Basic Prompting:]This approach organizes prompts into four components, as shown in Figure~\ref{P:method_1_2_4}(a): \textit{Task}—defines the problem and provides background information; \textit{Database Schema}—details the database structure; \textit{SQL}—presents two queries for equivalence evaluation; and \textit{Answer}—indicates the end of the prompt and where the LLM should start generating the response. Addressing SQL equivalence as a binary task (\textit{``Equivalent''} or \textit{``Non-Equivalent''}).
    \item[P2. Chain-of-Thought (CoT) Prompting:] Inspired by human problem-solving, this strategy breaks the task into intermediate reasoning steps~\citep{Wei2022ChainOT}. Extending basic prompting, it introduces a \textit{Steps} component, which explains query functionality and assesses equivalence through intermediate steps. Unlike Zero-Shot CoT prompting by~\citep{Kojima2022LargeLM}, which uses dynamic reasoning phrases, our approach (Figure~\ref{P:method_1_2_4}(b)) explicitly structures these steps in the prompt and forces LLMs to generate output based on the steps mentioned in the prompt.
    \item[P3. Few-Shot Prompting:] This method integrates task-specific examples (aka in-context examples) to improve LLMs performance~\citep{Brown2020LanguageMA}. For the experimentation, we used four in-context examples -- two equivalent and two non-equivalent examples. The ground truth explanations for these equivalent and non-equivalent examples are generated using \gpt-4 (Figure~\ref{P:method_1_2_4}(c)).  
\end{inparadesc}


\subsection{Multi-Stage Prompting}
\label{multi_stage_prompting} 

\begin{figure}
\begin{subfigure}[t]{0.3\textwidth}
    \centering
    \includegraphics[width=\linewidth]{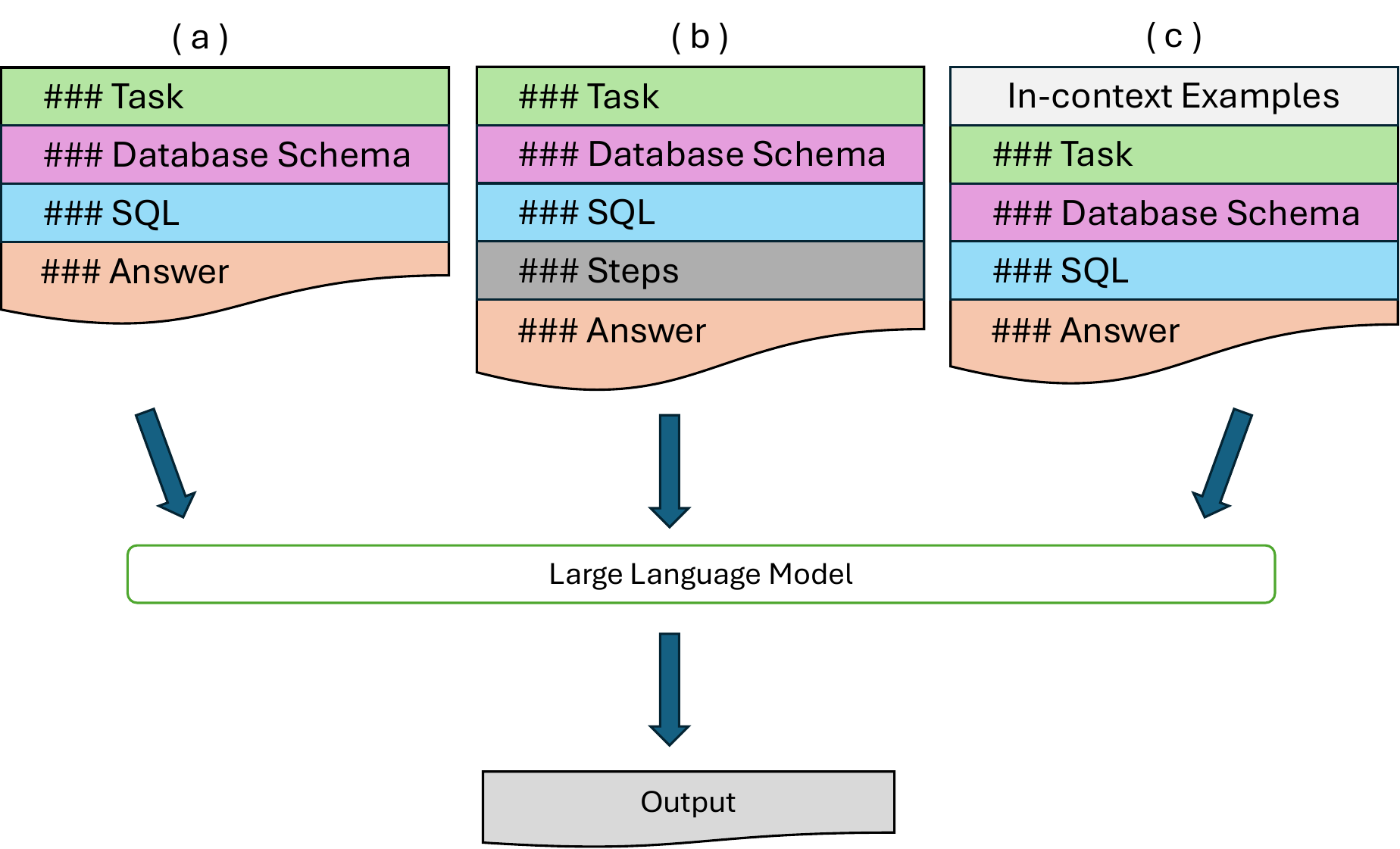}
    \caption{Single-stage prompting strategy, where (a) is the \textit{basic prompting}, (b) is the \textit{chain-of-thought prompting}, and (c) is the \textit{few-shot prompting}.}
    \label{P:method_1_2_4}
\end{subfigure}\hfill
\begin{subfigure}[t]{0.6\textwidth}
    \centering
    \includegraphics[width=1.0\textwidth]{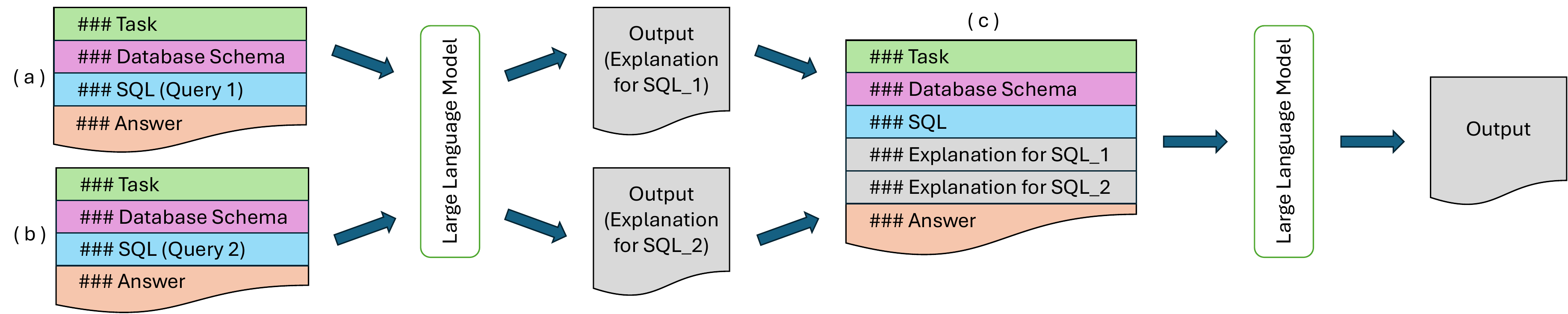}
    \caption{Multi-stage chain-of-thought prompting, where (a) and (b) show prompt construction for stage-1, and (c) shows the prompt construction at stage-2 of the pipeline.}
    \label{P:method_3}
\end{subfigure}
\caption{Prompting strategies employed (See Section~\ref{sec:proposed_model} for details).}
\end{figure}

As the name suggests, the multi-stage prompting strategy divides the reasoning process into multiple stages. Each stage uses the previous stage's output to formulate a new prompt, gradually building towards the final solution. Figure \ref{P:method_3} illustrates a multi-stage prompting strategy: \begin{inparadesc}
    \item[P4.] \textbf{Multi-Stage Chain-of-Thought Prompting}: extends the Chain-of-Thought approach by dividing the process into two distinct stages, as illustrated in Figure~\ref{P:method_3}. In the first stage, the LLM is tasked with providing concise explanations of the given SQL queries, focusing on breaking down their functionality and logic to ensure a clear understanding of their operations (Figures~\ref{P:method_3}(a) and~\ref{P:method_3}(b)). In the second stage, these explanations are incorporated into a new prompt that instructs the LLM to evaluate the equivalence of the SQL queries based on the provided details (Figure~\ref{P:method_3}(c)). Using the detailed explanations of the initial step, this approach aims to enhance the LLM's ability to make informed and accurate equivalence estimations.
\end{inparadesc}

\textbf{Adding Logical Plan in Prompt:} The SQL logical plan provides a detailed representation of a database query, outlining the high-level operations and transformations required to access data. In this work, we used the unoptimized SQL logical plan for each query, added under the \textit{``SQL"} section of the prompt. These plans were generated using Apache Calcite\footnote{https://calcite.apache.org/}, a popular open-source robust query optimization framework. Not all queries in the dataset were syntactically correct. Apache Calcite could not produce a valid logical plan for queries containing syntax errors. In such cases, the placeholder ``ERROR WHILE GENERATING PLAN'' was used to signify that the query could not be processed due to syntactic issues.

\noindent
\textbf{Classifying Prompt}: The outputs produced by the prompting strategies outlined in Section \ref{sec:proposed_model} are subsequently processed by a classifying prompt, which categorizes the results into three labels: \textit{``Equivalent''}, \textit{``Non-Equivalent''}, and \textit{``Unknown''}. A detailed explanation of this classification prompt is provided in the Appendix~\ref{app:classifying_output}.


\section{Results and Analysis}
\label{sec:results_and_analysis}
In this section, we present the results of our empirical evaluation, which includes an in-depth study of the performance of various LLMs. We analyze the impact of different prompting techniques and compare the results obtained using these LLMs with the state-of-the-art traditional approaches for SQL equivalence-checking tasks.

\subsection{Performance of LLMs on Benchmark Datasets}
\label{performance_of_llm}
We empirically evaluated various prompting strategies across different LLMs ---ranging from small scale \codellama-7B models to large \gpt-4 and \geminipro models ---over the three benchmark datasets for SQL equivalence. We present results separately for equivalent (EQ) query pairs and non-equivalent (NEQ) query pairs in each benchmark, as well as a geometric mean (GM) of both. Table~\ref{T:all} summarizes our results. 
We make the following important observations based on these results: First of all, nearly all LLMs perform well on equivalent query pairs, but struggle significantly with non-equivalent sets. This seems to suggest the presence of an inherent bias towards equivalence predictions, perhaps due to a similar bias in its training data. Only exception is \gpt-4 which is able to handle NEQ pairs, even with simple prompting strategies, quite well. Surprisingly and somewhat counter-intuitively, \gpt-4 performs poorly on the \calcite dataset, much worse than significantly smaller models such as \codellama-7B. 

\begin{wraptable}{r}{0.6\textwidth}
    \centering
    \resizebox{0.6\textwidth}{!}{
    \begin{tabular}{ll c ccc ccc}
    \toprule
    \multirow{2}{*}{\textbf{Methods}} & \multirow{2}{*}{\textbf{Models}} & \multicolumn{1}{c}{\textbf{Calcite}} & \multicolumn{3}{c}{\textbf{Spider+DIN}} &  \multicolumn{3}{c}{\textbf{SQLEquiQuest}}\\
    \cmidrule(lr){3-3}\cmidrule(lr){4-6}\cmidrule(lr){7-9}
    && EQ & EQ & NEQ & GM & EQ & NEQ &  GM \\
    \multicolumn{2}{c}{\textit{Ground truth}}& 232  & 385 & 189 &  & 307 & 192 & \\
    \midrule
    & \codellama-7B                 & 0.759 & 0.888 & 0.032 &  0.1686 &  0.993 & 0.000 & 0.000 \\
    Basic & \codellama-13B          & \underline{\textbf{0.991}} & 1.000 & 0.005 &  0.0707 & 0.997 & 0.057 & 0.238 \\
    Prompting & \gpt-3.5            & 0.819 & 0.948 & 0.153 & 0.3808 & 0.987 & 0.120 & 0.344 \\
    & \gpt-4                        & 0.690 & 0.873 & 0.667 &  0.7631 & 0.710 & 0.458 & \underline{\textbf{0.570}}\\
    & \geminipro                   & 0.931 & 0.982 & 0.032 &  0.1773 & 0.984 & 0.083 & 0.286\\
    \cmidrule(lr){1-9}
    
    & \codellama-7B                 & 0.388 & 0.966 & 0.069 & 0.2582 & 0.899 & 0.099 & 0.298\\ 
    Chain of & \codellama-13B       & 0.336 & 0.974 & 0.069 & 0.2592 & 0.98 & 0.010 & 0.099\\
    Thought & \gpt-3.5              & 0.625 & 0.912 & 0.291 & 0.5152 & 0.974 & 0.266 & 0.509\\
    Prompting & \gpt-4              & 0.668 & 0.875 & 0.714 &  \underline{\textbf{0.7904}} & 0.664 & 0.375 & 0.499\\
    & \geminipro                    & 0.875 & 0.964 & 0.138 & 0.3647 & 0.987 & 0.083 & 0.286\\
    \cmidrule(lr){1-9}
    
    Multi-Stage & \codellama-7B     & 0.651 & 0.883 & 0.026 & 0.1515 & 0.853 & 0.146 & 0.353\\ 
    Chain of & \codellama-13B       & \underline{\textit{0.970}} & 1.000 & 0.005 & 0.0707 & 0.837 & 0.005 & 0.065\\
    Thought & \gpt-3.5              & 0.728 & 0.953 & 0.175 & 0.4084 & 0.993 & 0.229 & 0.477\\
    Prompting & \gpt-4              & 0.694 & 0.917 & 0.571 & 0.7236 & 0.691 & 0.401 & 0.526\\ 
    & \geminipro                    & 0.858 & 0.987 & 0.063 & 0.2494 & 0.993 & 0.036 & 0.189\\
    \cmidrule(lr){1-9}
    
    & \codellama-7B                 & 0.966 & 0.862 & 0.302 & 0.5102 & 0.945 & 0.026 & 0.157\\
    Few-Shot & \codellama-13B       & 0.948 & 0.966 & 0.190 & 0.4284 & 0.974 & 0.042 & 0.202\\ 
    Prompting & \gpt-3.5            & 0.586 & 0.800 & 0.566 & 0.6729 & 0.964 & 0.208 & 0.448\\ 
    & \gpt-4                        & 0.651 & 0.919 & 0.672 & \underline{\textit{0.7859}} & 0.792 & 0.365 & \underline{\textit{0.538}}\\
    & \geminipro                    & 0.556 & 0.906 & 0.328 & 0.5451 & 0.958 & 0.141 & 0.368\\
    \bottomrule
    \end{tabular}
    }
    \caption{The performance of various LLMs under different prompting methods on \calcite, \spiderdin, and \exams dataset. Here, EQ refers to \textit{Equivalent Queries}, NEQ is \textit{Non-Equivalent Queries}, and GM is \textit{Geometric Mean} (best results are highlighted in bold, while the second-best results are indicated in italics). }
    \label{T:all}
\end{wraptable}


Secondly, it is important to note that the performance of all approaches declines for the \exams dataset. To understand their performance differences better, we present the performance on individual questions from the \exams dataset in Table~\ref{T:exams_iitd}. For complex and syntactically diverse queries in question 5, \gpt-3.5 with CoT prompting surpasses \gpt-4. Our analysis indicates that for Questions 1 to 4, the majority of the submissions done by students are syntactically similar to each other with minor mistakes, which leads to the non-equivalence of those query pairs. In contrast, Question 5 is significantly more complex, with student responses demonstrating substantial semantic variation. These findings suggest that \gpt-4 emphasizes semantic similarity, whereas \gpt-3.5 prioritizes syntactic structure, leading to improved performance for identifying non-equivalent query pairs for question 5.

\begin{table}[b]
    \resizebox{\textwidth}{!}{
    \begin{tabular}{ll ccc ccc ccc ccc ccc}
    \toprule
    \multirow{2}{*}{\textbf{Methods}} & \multirow{2}{*}{\textbf{Models}} & \multicolumn{3}{c}{\textbf{Question 1}} & \multicolumn{3}{c}{\textbf{Question 2}} & \multicolumn{3}{c}{\textbf{Question 3}} & \multicolumn{3}{c}{\textbf{Question 4}} & \multicolumn{3}{c}{\textbf{Question 5}}\\
    \cmidrule(lr){3-5}\cmidrule(lr){6-8}\cmidrule(lr){9-11}\cmidrule(lr){12-14}\cmidrule(lr){15-17}
    && EQ & NEQ & GM & EQ & NEQ & GM & EQ & NEQ & GM & EQ & NEQ & GM & EQ & NEQ & GM\\
    \multicolumn{2}{c}{\textit{Ground truth}}& 102 & 8 &  & 102 & 7 &  & 77 & 32 &  & 16 & 86 &  & 10 & 59 & \\
    \midrule
    
    & \codellama-7B        & \underline{\textbf{1.000}} & 0.000 & 0.0000 & \underline{\textit{0.980}} & 0.000 & 0.0000 & \underline{\textbf{1.000}} & 0.000 & 0.0000 & \underline{\textbf{1.000}} & 0.000 & 0.0000 & \underline{\textbf{1.000}} & 0.000 & 0.0000 \\
    Basic & \codellama-13B & \underline{\textbf{1.000}} & 0.000 & 0.0000 & \underline{\textbf{1.000}} & 0.143 & 0.3782 & \underline{\textbf{1.000}} & 0.000 & 0.0000 & \underline{\textbf{1.000}} & 0.000 & 0.0000 & \underline{\textit{0.900}} & 0.169 & 0.3900 \\
    Prompting & \gpt-3.5   & \underline{\textit{0.990}} & 0.250 & 0.4975 & \underline{\textbf{1.000}} & 0.143 & 0.3782 & \underline{\textbf{1.000}} & 0.156 & 0.3950 & \underline{\textbf{1.000}} & 0.070 & 0.2646 & 0.700 & 0.153 & 0.3273 \\
    & \gpt-4               & 0.667 & \underline{\textbf{1.000}} & \underline{\textit{0.8167}} & 0.637 & \underline{\textit{0.857}} & 0.7389 & 0.779 & \underline{\textit{0.531}} & \underline{\textit{0.6432}} & \underline{\textbf{1.000}} & \underline{\textbf{0.500}} & \underline{\textbf{0.7071}} & \underline{\textit{0.900}} & 0.237 & 0.4618 \\
    & \geminipro           & 0.961 & 0.125 & 0.3466 & \underline{\textbf{1.000}} & 0.286 & 0.5348 & \underline{\textbf{1.000}} & 0.094 & 0.3066 & \underline{\textbf{1.000}} & 0.047 & 0.2168 & \underline{\textit{0.900}} & 0.102 & 0.3030 \\
    \cmidrule(lr){1-17}
    
    & \codellama-7B           & 0.863 & 0.625 & 0.7344 & 0.931 & 0.286 & 0.5160 & 0.974 & 0.063 & 0.2477 & \underline{\textit{0.938}} & 0.023 & 0.1469 & 0.300 & 0.136 & 0.2020 \\
    Chain of & \codellama-13B & 0.980 & 0.125 & 0.3500 & 0.961 & 0.000 & 0.0000 & \underline{\textbf{1.000}} & 0.000 & 0.0000 & \underline{\textbf{1.000}} & 0.000 & 0.0000 & \underline{\textbf{1.000}} & 0.017 & 0.1304 \\
    Thought & \gpt-3.5        & \underline{\textit{0.990}} & 0.250 & 0.4975 & \underline{\textit{0.980}} & 0.143 & 0.3744 & \underline{\textbf{1.000}} & 0.094 & 0.3066 & \underline{\textbf{1.000}} & 0.128 & 0.3578 & 0.500 & \underline{\textbf{0.576}} & \underline{\textit{0.5367}} \\
    Prompting & \gpt-4        & 0.559 & \underline{\textbf{1.000}} & 0.7477 & 0.676 & \underline{\textbf{1.000}} & \underline{\textbf{0.8222}} & 0.727 & 0.500 & 0.6029 & 0.875 & \underline{\textit{0.360}} & 0.5612 & 0.800 & 0.169 & 0.3677 \\
    & \geminipro              & 0.980 & 0.125 & 0.3500 & \underline{\textbf{1.000}} & 0.143 & 0.3782 & \underline{\textbf{1.000}} & 0.063 & 0.2510 & \underline{\textit{0.938}} & 0.070 & 0.2562 & \underline{\textit{0.900}} & 0.102 & 0.3030 \\
    \cmidrule(lr){1-17}
    
    Multi-Stage & \codellama-7B & 0.882 & 0.250 & 0.4696 & 0.931 & 0.143 & 0.3649 & 0.779 & 0.063 & 0.2215 & 0.625 & 0.128 & 0.2828 & 0.700 & 0.203 & 0.3770 \\
    Chain of & \codellama-13B   & \underline{\textbf{1.000}} & 0.000 & 0.0000 & 0.539 & 0.000 & 0.0000 & \underline{\textbf{1.000}} & 0.031 & 0.1761 & 0.813 & 0.000 & 0.0000 & \underline{\textbf{1.000}} & 0.000 & 0.0000 \\
    Thought & \gpt-3.5          & \underline{\textbf{1.000}} & 0.250 & 0.5000 & \underline{\textbf{1.000}} & 0.143 & 0.3782 & \underline{\textbf{1.000}} & 0.125 & 0.3536 & \underline{\textbf{1.000}} & 0.163 & 0.4037 & 0.800 & \underline{\textit{0.390}} & \underline{\textbf{0.5586}} \\
    Prompting & \gpt-4          & 0.627 & \underline{\textbf{1.000}} & 0.7918 & 0.725 & 0.714 & 0.7195 & 0.688 & \underline{\textbf{0.656}} & \underline{\textbf{0.6718}} & \underline{\textit{0.938}} & \underline{\textit{0.360}} & \underline{\textit{0.5811}} & 0.600 & 0.203 & 0.3490 \\
    & \geminipro                & \underline{\textbf{1.000}} & 0.125 & 0.3536 & \underline{\textbf{1.000}} & 0.143 & 0.3782 & 0.974 & 0.031 & 0.1738 & \underline{\textbf{1.000}} & 0.012 & 0.1095 & \underline{\textbf{1.000}} & 0.051 & 0.2258 \\
    \cmidrule(lr){1-17}
    
    & \codellama-7B           & 0.960 & 0.000 & 0.0000 & \underline{\textbf{1.000}} & 0.000 & 0.0000 & 0.922 & 0.031 & 0.1691 & 0.625 & 0.000 & 0.0000 & \underline{\textit{0.900}} & 0.068 & 0.2474 \\
    Few-Shot & \codellama-13B & \underline{\textbf{1.000}} & 0.000 & 0.0000 & \underline{\textbf{1.000}} & 0.000 & 0.0000 & 0.961 & 0.156 & 0.3872 & \underline{\textbf{1.000}} & 0.023 & 0.1517 & 0.500 & 0.017 & 0.0922 \\
    Prompting & \gpt-3.5      & 0.980 & 0.375 & 0.6062 & 0.961 & 0.286 & 0.5243 & \underline{\textit{0.987}} & 0.125 & 0.3512 & \underline{\textit{0.938}} & 0.198 & 0.4310 & 0.700 & 0.237 & 0.4073 \\
    & \gpt-4                  & 0.755 & \underline{\textit{0.875}} & 0.8128 & 0.686 & \underline{\textit{0.857}} & \underline{\textit{0.7667}} & 0.922 & 0.438 & 0.6355 & \underline{\textit{0.938}} & 0.349 & 0.5722 & \underline{\textbf{1.000}} & 0.220 & 0.4690 \\
    & \geminipro              & 0.941 & 0.750 & \underline{\textbf{0.8401}} & \underline{\textit{0.980}} & 0.143 & 0.3744 & 0.961 & 0.188 & 0.4251 & \underline{\textit{0.938}} & 0.070 & 0.2562 & \underline{\textit{0.900}} & 0.135 & 0.3486 \\
    \bottomrule
    \end{tabular}
    }
    \caption{Performance of LLMs on \exams dataset. 
    (The best results are highlighted in bold, while the second-best results are indicated in italics)}
    \label{T:exams_iitd}
\end{table}

\begin{figure}[t]
    \centering
    \includegraphics[width=1.0\textwidth]{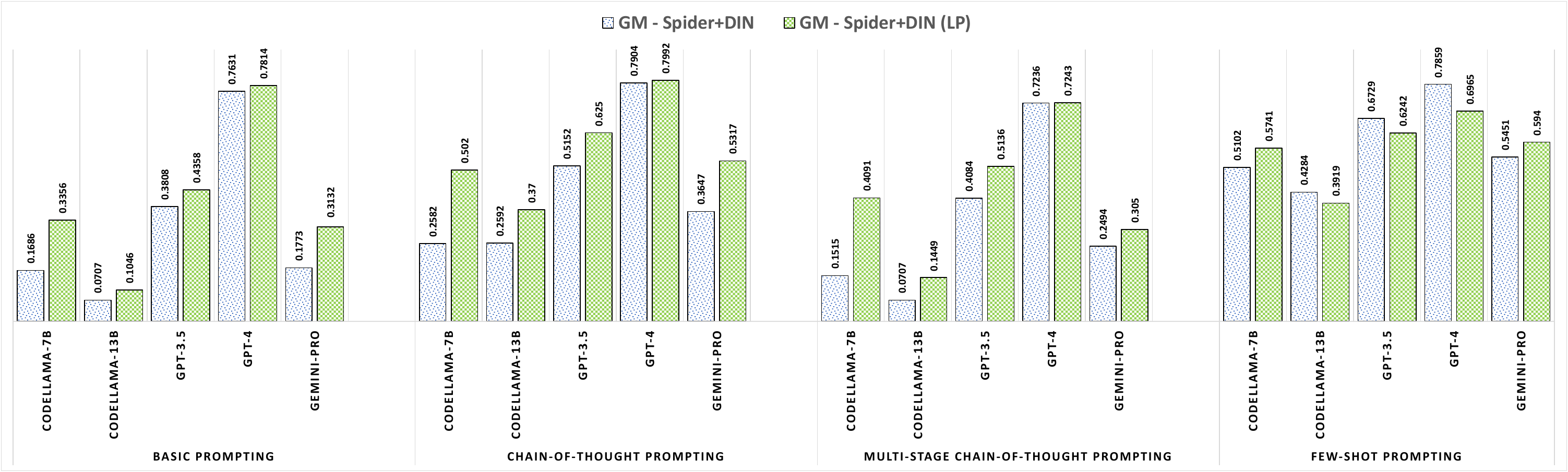}
    \caption{Performance of different LLMs under different prompting methods on \spiderdin; with and without adding a logical plan (LP) in the prompt. GM refers to Geometric Mean.}
    \label{Fig:comp_spider_din_lp}
\end{figure}

We explored a novel approach to SQL query equivalence checking by incorporating the query logical plan (LP) into the prompt. The results on the \spiderdin benchmark, presented in Figure~\ref{Fig:comp_spider_din_lp}, demonstrate that the inclusion of LP consistently enhances performance across various prompting methods, with the exception of the few-shot setting. Notably, substantial performance gains were observed for the smallest model, \codellama-7B. The minimal improvement in few-shot settings may be attributed to suboptimal example selection. Additionally, we fine-tuned the publicly available \codellama-13B model for the SQL query equivalence classification task (refer to Appendix Section~\ref{app:finetuneing}). Fine-tuning significantly improved the model’s accuracy in identifying non-equivalent SQL query pairs, while performance on equivalent pairs remained stable, with a slight decrease.

\noindent
\textbf{Note:} Due to budget constraints, we were unable to test the performance of incorporating the query logical plan in the prompt for other datasets.

These findings provide valuable insights into the strengths and weaknesses of different LLMs, particularly in handling queries they are unlikely to have encountered during training. Also, we noticed that the explanations from the \gpt \space family of LLMs tend to provide good guidance for understanding and revising the SQL queries (for example, to help students revise their SQL).

\subsection{Comparison with Traditional Approaches for SQL Equivalence Checking}
\label{comparision_with_traditional_approaches}

Due to its importance in query-optimization and reuse of query results in cloud platforms, SQL equivalence checking is extensively studied, resulting in sophisticated tools such as \verieql~\citep{He2024VeriEQLBE}, \sqlsolver~\citep{Ding2023ProvingQE}, SPES~\citep{Zhou2022SPESAS}, UDP~\citep{UDP}, and EQUITAS~\citep{Zhou2019AutomatedVO}. We summarize the performance of these tools in terms of their coverage by presenting the number of `Supported' and `Unsupported' queries for each. `Supported' queries are those the models can verify, while `Unsupported' queries fall outside their scope. As these numbers show, the best performing tool \sqlsolver can only support $71\%$ of \spiderdin and a mere $30\%$ of queries of \exams benchmarks. In comparison, \gpt-4 scores $80\%$ in \spiderdin and $60\%$ in \exams in its correct predictions.


\begin{wraptable}{r}{0.55\textwidth}
    \centering
    \resizebox{0.55\textwidth}{!}{
    \begin{tabular}{ccll}
        \toprule
        \textbf{Method} & \textbf{Dataset} & \textbf{Supported} & \textbf{Unsupported} \\
        \midrule
          & \calcite & 232  \footnotesize{{(\textcolor{violet}{161})}}& 0 \footnotesize{{(\textcolor{violet}{0})}}\\
          \sqlsolver~\citep{Ding2023ProvingQE} & \spiderdin & 412  \footnotesize{(\textcolor{violet}{375})} & 162 \footnotesize{{(\textcolor{violet}{86})}}\\
          & \exams & 151  \footnotesize{(\textcolor{violet}{124})} & 348 \footnotesize{{(\textcolor{violet}{182})}}\\
         \midrule
          & \calcite & 119 \footnotesize{(\textcolor{violet}{86})} & 113 \footnotesize{{(\textcolor{violet}{75})}}\\
         \verieql~\citep{He2024VeriEQLBE} & \spiderdin & 206   \footnotesize{(\textcolor{violet}{182})} & 368 \footnotesize{{(\textcolor{violet}{273})}} \\
          & \exams & 14   \footnotesize{(\textcolor{violet}{13})} & 485 \footnotesize{{(\textcolor{violet}{293})}}\\
        \midrule
        SPES$^*$~\citep{Zhou2022SPESAS} & \calcite & 95  & 137 \\
        \midrule
        EQUITAS$^*$~\citep{Zhou2019AutomatedVO} & \calcite & 67  & 165 \\
        \midrule
        UDP$^*$~\citep{UDP} & \calcite & 33  &  199\\
        \bottomrule
    \end{tabular}
    }
    \caption{Comparison of coverage of queries using formal methods. Results marked with `$^*$' are sourced from the original research papers. The numbers in parentheses indicate the overlap with correct results from \gpt-4. \\ }
    \label{tab:comp_trad}
\end{wraptable}

Furthermore, the table also highlights the advantage of using LLMs for this task. For instance, of the 151 query-pairs in \exams that \sqlsolver supports, \gpt-4 can correctly classify more than 80\% of them. Even more notable is its ability to correctly answer $52\%$ of the query-pairs \emph{unsupported} by the best performing \sqlsolver. Similar behavior can be observed with other datasets and other SQL equivalence checkers based on formal methods. Although it is not apparent from these results, it is worth noting that each of the compared tools exhibit certain common failings repeatedly -- e.g., \verieql is case-sensitive and prone to misclassification, while \sqlsolver struggling with column permutation and subset relationships. The LLM-based approach we explored in this paper overcomes these issues.


\subsection{Impact of Query Complexity} 
\label{performance_of_pre_trained_llms_against_query_complexity}
        
We examined the impact of query complexity on the performance of LLMs using the \spiderdin dataset, categorizing queries based on complexity levels (Easy, Medium, Hard, Extra-Hard) as defined by the Spider dataset authors~\citep{test-suite-sql-eval}. Our findings indicate that \codellama-7B and \codellama-13B perform well on equivalent queries across varying levels of SQL query complexity but struggle with non-equivalent queries, consistent with their bias observed in Section~\ref{performance_of_llm}.  Notably, incorporating in-context examples significantly improves performance on non-equivalent queries for all models except \gpt-4. In contrast, \gpt-4 maintains consistent performance across various prompting strategies and complexity levels, highlighting its superior reasoning capabilities. 

\begin{figure}[t]
    \centering
    \includegraphics[width=1.0\textwidth]{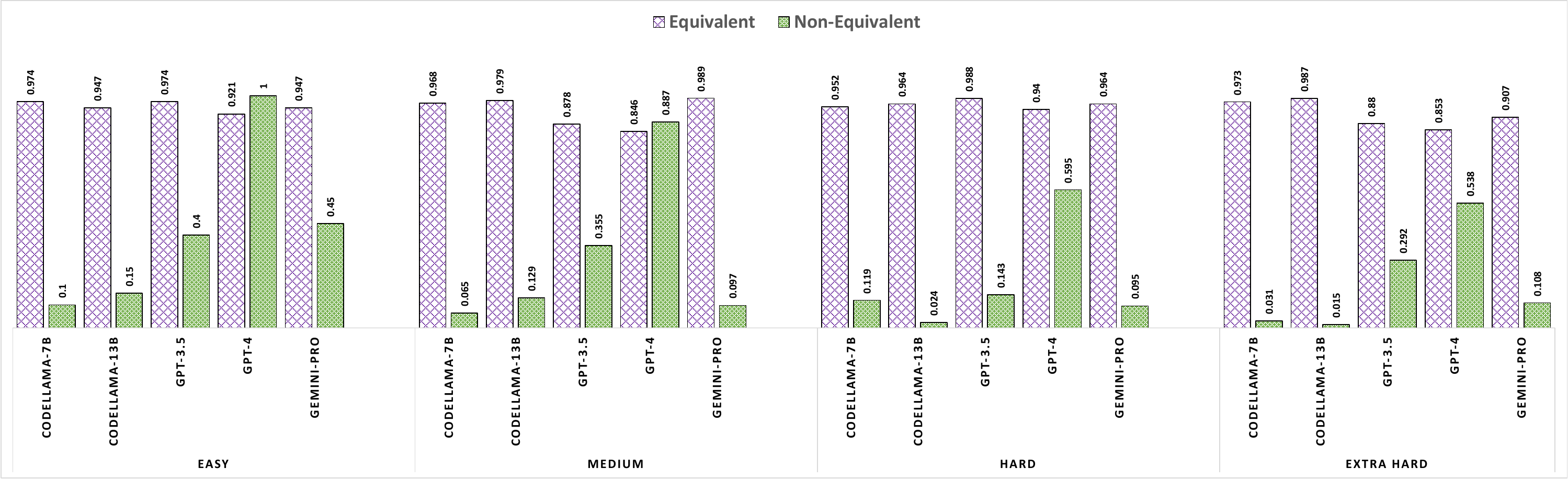}
    \caption{Performance of \gpt-4 on \spiderdin dataset split based on difficulty level.  }
    \label{Fig:spider_difficulty}
\end{figure}

Figure~\ref{Fig:spider_difficulty} presents model accuracy on \spiderdin dataset using CoT prompting. The results indicate a clear decline in performance on non-equivalent query pairs as complexity increases, whereas equivalent query pairs remain largely unaffected by increasing query difficulty. This trend persists across different prompting techniques.

\subsection{Challenges}
\label{sec:future_work}

The research encountered several notable challenges, particularly in the context of managing large language model (LLM) output. A key challenge was the sensitivity of LLM responses to prompt variations. Even minor modifications, such as altering a single character (e.g., adding a semicolon) or changing capitalization, could lead to significantly different output. Consequently, the construction of effective prompts required meticulous experimentation. All prompts presented and shared in this study were carefully designed and refined through iterative testing with each LLM used.

Furthermore, the inherent unpredictableness of LLM outputs posed significant difficulties. Although output constraints could be partially achieved by embedding specific instructions within the prompts, several issues persisted. These included the repetition of lines, the generation of contradictory statements, and the inclusion of extraneous ASCII characters. These challenges were particularly pronounced in smaller models, such as \codellama-7B and \codellama-13B. To address these issues, an auxiliary prompt (Refer Section ~\ref{app:classifying_output} in Appendix) was used. Repeated lines and related issues were managed by carefully pruning irrelevant content from the generated outputs.

\section{Conclusions \& Future Work}
\label{sec:conclusion}

\raj{This study examines the use of large language models (LLMs) to understand SQL queries and determine their equivalence. Results in Section~\ref{sec:results_and_analysis} reveal that, despite lacking theoretical guarantees, LLMs show significant potential in this task, handling queries beyond the scope of traditional methods. However, most models (except \gpt-4) exhibit a bias toward incorrectly labeling queries as equivalent, a limitation mitigated through fine-tuning.}

\raj{While formal proof of LLM reasoning remains elusive, their ability to provide intermediate explanations highlights their utility for non-critical applications, such as grading and feedback on SQL assignments. To fully realize their potential in critical contexts, further work is required on fine-tuning, managing query complexity, and refining equivalence definitions.}

\raj{Future research directions include leveraging Retrieval-Augmented Generation (RAG) to enhance in-context learning with similar query examples~\citep{symkgqa2024} and integrating LLMs with symbolic proof systems to better analyze the logical structure of complex SQL queries.}

\clearpage


\bibliography{colm2025_conference}
\bibliographystyle{colm2025_conference}

\newpage
\appendix
\textbf{\Large APPENDIX}

\setcounter{figure}{0}
\renewcommand\thefigure{\thesection.\arabic{figure}}

\setcounter{table}{0}
\renewcommand\thetable{\thesection.\arabic{table}} 

\setcounter{section}{0}
\renewcommand{\thesection}{\Alph{section}}

\section{Choice of Large Language Models (LLMs)}
\label{app:choice_of_llms}
The proliferation of LLMs makes evaluating every new model impossible. This study focuses on selected LLMs based on budget and computational constraints, with the methodology adaptable to other models. The insights gained are broadly applicable across various LLMs. Below is an overview of the selected LLMs:
    
\begin{itemize}

    \item \textbf{\codellama}~\cite{Rozire2023CodeLO}: Built on \llama 2 by Meta, \codellama models are designed for code synthesis and understanding, supporting stable outputs up to 100,000 tokens during inference. Variants of \codellama include ``Base'', ``Python'', and ``Instruct'' models, with parameter sizes of 7B, 13B, 34B, and 70B. We used the ``\codellama-7B-Instruct-hf'' and ``\codellama-13B-Instruct-hf'' models, optimized for instruction-following tasks.

    \item \textbf{\gpt} \cite{gpt,Achiam2023GPT4TR}: Stands for \texttt{\textbf{G}enerative} \texttt{\textbf{P}re-trained} \texttt{\textbf{T}ransformer}, is a family of decoder-only autoregressive large language models developed by OpenAI. For the scope of this work, we used ``gpt-3.5-turbo-0125'' (175B parameters) and ``gpt-4-0125-preview'' (~1.8T parameters, estimated\footnote{https://web.archive.org/web/20230712123915/https://the-decoder.com/gpt-4-architecture-datasets-costs-and-more-leaked/}) models. These models excel in natural language tasks and are accessed via API. For simplicity, this paper will refer to ``gpt-3.5-turbo-0125'' as ``\gpt-3.5'' and ``gpt-4-0125-preview'' as ``\gpt-4''.

    \item \textbf{\geminipro} \cite{Anil2023GeminiAF}: Developed by Google DeepMind, Gemini models are decoder-only architectures available via API. This study uses ``\geminipro'', known for its versatility in natural language understanding and generation tasks.
    
\end{itemize}

LLMs specialized in text-to-SQL tasks, such as SQLCoder, were excluded as they lacked flexibility in generating explanatory responses. Instead, this study focuses on models with broader adaptability across diverse prompts.

\section{Prompting Strategies}
\label{app:promting_strategies}
This section presents various state-of-the-art prompting strategies (with sample prompt) for Large Language Models (LLMs) that we have adopted for SQL equivalence checking.

\subsection{Basic Prompting}
\label{app:basic_prompting}
As the name suggests, the basic prompting strategy relies on a minimalistic prompt structure containing only essential information. This strategy comprises four key components, as illustrated in Figure~\ref{P:method_1_2_4}(a):\\
(i) \textbf{Task} -- Defines the problem and provides background information.\\
(ii) \textbf{Database Schema} -- Details the database structure.\\
(iii) \textbf{SQL} -- Provides SQL queries (with/without logical plan) for equivalence evaluation.\\
(iv) \textbf{Answer} -- Indicates the end of the prompt and where the LLM should start generating the response.\\
The prompt structure used for basic prompting is shown in Figure~\ref{app:P:basic_prompting}.

\begin{figure}[ht]
    \centering
    \begin{tcolorbox}[title=P1 - Basic Prompting]
        \small
        $\#\#\#$ Task\\
        Given the database schema and two SQL queries, SQL\_1 and SQL\_2, determine whether SQL\_1 and SQL\_2 are "Equivalent" or "Non Equivalent". Two SQL queries are "Equivalent" if both queries produce the same result when executed on the given database schema.
\end{tcolorbox}
\end{figure}

\begin{figure}[ht]
    \centering
    \begin{tcolorbox}[]
        \small
        $\#\#\#$ Database Schema\\
        The query will run on a database with the following schema: \\
        Table $<$table\_name$>$, columns = [ *, $<$column\_name$>$, $\hdots$, $<$column\_name$>$ ]\\
        \vdots \\
        Table $<$table\_name$>$, columns = [ *, $<$column\_name$>$, $\hdots$, $<$column\_name$>$ ]\\\\
        Foreign\_keys = [ $<$table\_name$>$.$<$column\_name$>$ = $<$table\_name$>$.$<$column\_name$>$, $\hdots$ ] \\
        Primary\_keys = [ $<$table\_name$>$.$<$column\_name$>$, $<$table\_name$>$.$<$column\_name$>$, $\hdots$ ]\\\\
        $\#\#\#$ SQL \\
        $[$SQL\_1$]$ $<$sql query$>$  \\
        $<$SQL\_1 Logical Plan$>$\\\\
        $[$SQL\_2$]$ $<$sql query$>$ \\
        $<$SQL\_2 Logical Plan$>$\\\\
        $\#\#\#$ Answer
    \end{tcolorbox}
    \caption{Prompt used for Basic Prompting.}
    \label{app:P:basic_prompting}
\end{figure}

\subsection{Chain-of-Thought (CoT) Prompting}
\label{app:chain_of_thought_prompting}
This prompting strategy compels the LLM to decompose the task into intermediate reasoning steps. These steps establish context and guide the LLM toward problem-solving. The strategy consists of five key components, as depicted in Figure~\ref{P:method_1_2_4}(b):\\
(i) \textbf{Task} -- Defines the problem and provides background information.\\
(ii) \textbf{Database Schema} -- Details the database structure.\\
(iii) \textbf{SQL} -- Provides SQL queries (with/without logical plan) for equivalence evaluation.\\
(iv) \textbf{Steps} -- Directs the prompt to explain the queries and evaluate their equivalence through intermediate steps.\\
(v) \textbf{Answer} -- Indicates the end of the prompt and where the LLM should start generating the response.\\
The prompt structure used for Chain-of-Thought (CoT) Prompting is shown in Figure~\ref{app:P:cot_prompting}.

\begin{figure}[ht]
\vspace{10px}
    \centering
    \begin{tcolorbox}[title=P2 - Chain-of-Thought (CoT) Prompting]
    \small
    $\#\#\#$ Task\\
    Given the database schema and two SQL queries, SQL\_1 and SQL\_2, determine whether SQL\_1 and SQL\_2 are "Equivalent" or "Non Equivalent". Two SQL queries are "Equivalent" if both queries produce the same result when executed on the given database schema.\\ \\
    $\#\#\#$ Database Schema\\
    The query will run on a database with the following schema: \\
    Table $<$table\_name$>$, columns = [ *, $<$column\_name$>$, $\hdots$, $<$column\_name$>$ ]\\
    \vdots \\
    Table $<$table\_name$>$, columns = [ *, $<$column\_name$>$, $\hdots$, $<$column\_name$>$ ]\\ \\
    Foreign\_keys = [ $<$table\_name$>$.$<$column\_name$>$ = $<$table\_name$>$.$<$column\_name$>$, $\hdots$ ] \\
    Primary\_keys = [ $<$table\_name$>$.$<$column\_name$>$, $<$table\_name$>$.$<$column\_name$>$, $\hdots$ ]\\ \\
    $\#\#\#$ SQL \\
    $[$SQL\_1$]$ $<$sql query$>$ \\
    $<$SQL\_1 Logical Plan$>$\\\\
    $[$SQL\_2$]$ $<$sql query$>$ \\ 
    $<$SQL\_2 Logical Plan$>$
    \end{tcolorbox}
\end{figure}

\begin{figure}[ht]
    \centering
    \begin{tcolorbox}
    \small
    
    $\#\#\#$ Steps\\
    Let's think step by step as follows: \\
    Step 1: Explain in brief each of the two queries, SQL\_1 and SQL\_2.\\
    Step 2: Based on the explanation, determine whether SQL\_1 and SQL\_2 are "Equivalent" or "Non Equivalent". \\
    Step 3: Provide the analysis and reasoning for the conclusion. \\ \\
    $\#\#\#$ Answer
    \end{tcolorbox}
    \caption{Prompt used for Chain-of-Thought (CoT) Prompting.}
    \label{app:P:cot_prompting}
\end{figure}

\subsection{Few-Shot Prompting}
\label{app:few_shot_prompting}
This prompting strategy incorporates task-specific examples within the prompt. These examples assist LLMs in understanding the problem-solving approach, ultimately enhancing their overall performance. By providing concrete examples, the model gains insight into expected patterns, logical structures, and solution methodologies. This approach helps reduce ambiguity, improves response accuracy, and facilitates better generalization to similar tasks. For the experimentation, we used four in-context examples -- two equivalent and two non-equivalent. These examples were randomly sampled from the dataset and then fixed for all test samples to ensure consistency. The ground truth explanations for these examples are generated using \gpt-4. The prompt strategy consist of five key components, as depicted in Figure~\ref{P:method_1_2_4}(c):\\
(i) \textbf{Examples} -- Provides the in-context examples.\\
(ii) \textbf{Task} -- Defines the problem and provides background information.\\
(iii) \textbf{Database Schema} -- Details the database structure.\\
(iv) \textbf{SQL} -- Provides SQL queries (with/without logical plan) for equivalence evaluation.\\
(v) \textbf{Answer} -- Indicates the end of the prompt and where the LLM should start generating the response.\\
The prompt structure used for Few-Shot Prompting is shown in Figure~\ref{app:P:few_shot}.

\begin{figure}[!ht]
    \vspace{10px}
    \centering
    \begin{tcolorbox}[title=P3 - Few-Shot Prompting]
    \small
    $\#\#\#$ Example 1 \\
    \vdots \\
    $\#\#\#$ Example 2 \\
    \vdots \\
    $\#\#\#$ Example 3 \\
    \vdots \\
    $\#\#\#$ Example 4 \\
    \vdots \\
    $\#\#\#$ Question \\
    $\#\#\#$ Task\\
    Given the database schema and two SQL queries, SQL\_1 and SQL\_2, determine whether SQL\_1 and SQL\_2 are "Equivalent" or "Non Equivalent". Two SQL queries are "Equivalent" if both queries produce the same result when executed on the given database schema.\\ \\
    $\#\#\#$ Database Schema\\
    The query will run on a database with the following schema: \\
    Table $<$table\_name$>$, columns = [ *, $<$column\_name$>$, $\hdots$, $<$column\_name$>$ ]\\
    \vdots \\
    Table $<$table\_name\_t$>$, columns = [ *, $<$column\_name$>$, $\hdots$, $<$column\_name$>$ ]\\ \\
    Foreign\_keys = [ $<$table\_name$>$.$<$column\_name$>$ = $<$table\_name$>$.$<$column\_name$>$, $\hdots$ ] \\
    Primary\_keys = [ $<$table\_name$>$.$<$column\_name$>$, $<$table\_name$>$.$<$column\_name$>$, $\hdots$ ]
    \end{tcolorbox}
\end{figure}

\begin{figure}[!ht]
    \centering
    \begin{tcolorbox}
    \small
    $\#\#\#$ SQL \\
    $[$SQL\_1$]$ $<$sql query$>$  \\
    $<$SQL\_1 Logical Plan$>$\\ \\
    $[$SQL\_2$]$ $<$sql query$>$ \\
    $<$SQL\_1 Logical Plan$>$\\ \\
    $\#\#\#$ Answer
    \end{tcolorbox}
    \caption{Prompt used in Few-Shot Prompting.}
    \label{app:P:few_shot}
\end{figure}

\subsection{Multi-Stage Chain-of-Thought Prompting}
\label{app:Multistage_cot_prompting}
Multi-stage CoT divides the process into two distinct stages (Figure~\ref{P:method_3}). In the first stage, the LLM is tasked with providing concise explanations of the given SQL queries. The prompt used at this stage consists of four key components (Figure~\ref{app:P:multi_cot_prompting_a}):\\
(i) \textbf{Task} -- Defines the problem and provides background information. (Note: For this prompt, the task is to generate a brief description of the given SQL query)\\
(ii) \textbf{Database Schema} -- Details the database structure.\\
(iii) \textbf{SQL} -- Provides SQL queries (with/without a logical plan) for which an explanation needs to be generated.\\
(iv) \textbf{Answer} -- Indicates the end of the prompt and where the LLM should start generating the response.

\begin{figure}[!ht]
    \vspace{10px}
    \centering
    \begin{tcolorbox}[title=P4 - Prompt for explaining $<$SQL\_1 / SQL\_2$>$]
    \small
    $\#\#\#$ Task\\
    Given the database schema and an SQL query, i.e., $<$ SQL\_1 / SQL\_2$>$, briefly describe the SQL query.\\ \\
    $\#\#\#$ Database Schema\\
    The query will run on a database with the following schema: \\
    Table $<$table\_name$>$, columns = [ *, $<$column\_name$>$, $\hdots$, $<$column\_name$>$ ]\\
    \vdots \\
    Table $<$table\_name$>$, columns = [ *, $<$column\_name$>$, $\hdots$, $<$column\_name$>$ ]\\ \\
    Foreign\_keys = [ $<$table\_name$>$.$<$column\_name$>$ = $<$table\_name$>$.$<$column\_name$>$, $\hdots$ ] \\
    Primary\_keys = [ $<$table\_name$>$.$<$column\_name$>$, $<$table\_name$>$.$<$column\_name$>$, $\hdots$ ]\\ \\
    $\#\#\#$ SQL \\
    $[$SQL\_1 / SQL\_2$]$ $<$sql query$>$ \\
    $<$SQL\_1/SQL\_2 Logical Plan$>$\\ \\
    $\#\#\#$ Answer
    \end{tcolorbox}
    \caption{Prompt used in Multi-Stage Chain-of-Thought Prompting for explaining SQL queries.}
    \label{app:P:multi_cot_prompting_a}
\end{figure}

Subsequently, the descriptions generated for the SQL queries in stage one are used to construct the prompt for stage two (Figure~\ref{app:P:multi_cot_prompting_b}). The stage two prompt consists of six key components:\\
(i) \textbf{Task} -- Defines the problem and provides background information.\\
(ii) \textbf{Database Schema} -- Details the database structure.\\
(iii) \textbf{SQL} -- Provides SQL queries (with/without a logical plan) for equivalence evaluation.\\
(iv) \textbf{Explanation for SQL\_1} -- Provides the description about SQL\_1 obtained from stage 1.\\
(v) \textbf{Explanation for SQL\_2} -- Provides the description about SQL\_2 obtained from stage 1.\\
(vi) \textbf{Answer} -- Indicates the end of the prompt and where the LLM should start generating the response.

\begin{figure}[!ht]
    \vspace{10px}
    \centering
    \begin{tcolorbox}[title=P4 - Prompt for checking query equivalence]
    \small
    $\#\#\#$ Task\\
    Given the database schema and two SQL queries with explanation, SQL\_1 and SQL\_2, determine whether SQL\_1 and SQL\_2 are "Equivalent" or "Non Equivalent". Two SQL queries are "Equivalent" if both queries produce the same result when executed on the given database schema.\\ \\
    $\#\#\#$ Database Schema\\
    The query will run on a database with the following schema: \\
    Table $<$table\_name$>$, columns = [ *, $<$column\_name$>$, $\hdots$, $<$column\_name$>$ ]\\
    \vdots \\
    Table $<$table\_name\_t$>$, columns = [ *, $<$column\_name$>$, $\hdots$, $<$column\_name$>$ ]\\ \\
    Foreign\_keys = [ $<$table\_name$>$.$<$column\_name$>$ = $<$table\_name$>$.$<$column\_name$>$, $\hdots$ ] \\
    Primary\_keys = [ $<$table\_name$>$.$<$column\_name$>$, $<$table\_name$>$.$<$column\_name$>$, $\hdots$ ]\\ \\
    $\#\#\#$ SQL \\
    $[$SQL\_1$]$ $<$sql query$>$ \\
    $<$SQL\_1 Logical Plan$>$\\ \\
    $[$SQL\_2$]$ $<$sql query$>$ \\
    $<$SQL\_2 Logical Plan$>$\\ \\
    $\#\#\#$ Explanation for SQL\_1\\
    $<$Explanation$>$\\ \\
    $\#\#\#$ Explanation for SQL\_2\\
    $<$Explanation$>$\\ \\
    $\#\#\#$ Answer \\
    \end{tcolorbox}
    \caption{Prompt used in Multi-Stage Chain-of-Thought Prompting for checking SQL query equivalence.}
    \label{app:P:multi_cot_prompting_b}
\end{figure}

\subsection{Classifying Prompt}
\label{app:classifying_output}
It is important to note that many LLMs (with the notable exception of the \gpt \space family) tend to generate large amounts of text in response to prompting strategies, even when not requested. These responses often include explanations, repetitive statements, and counterexamples generated by the LLM to help reach its conclusions. Although this verbosity may be acceptable for a human judge, it becomes problematic when we want a machine to process these responses and make a decision. Furthermore, manually evaluating all the responses will be time-consuming and expensive.

There are two commonly used techniques for identifying the final result of these verbose responses. The first technique involves crafting the prompt in such a way that the LLM is forced to generate a standardized, templated response that can be programmatically parsed \cite{Liu2023GEvalNE}. This can be achieved by detecting a specific textual pattern, as described in resources such as the LangChain output parser, designed to convert the output of LLMs into a certain format.

The second technique involves a classifier that takes in the response text to produce the required (boolean) answer. However, some LLMs, such as \codellama, might deviate from the expected response format even when the former approach is used. On the other hand, the latter approach is challenging due to the need to train a high-accuracy classifier capable of interpreting the diverse and verbose outputs of the LLM. 

\begin{figure}[tp]
    \centering
    \includegraphics[width=0.65\textwidth]{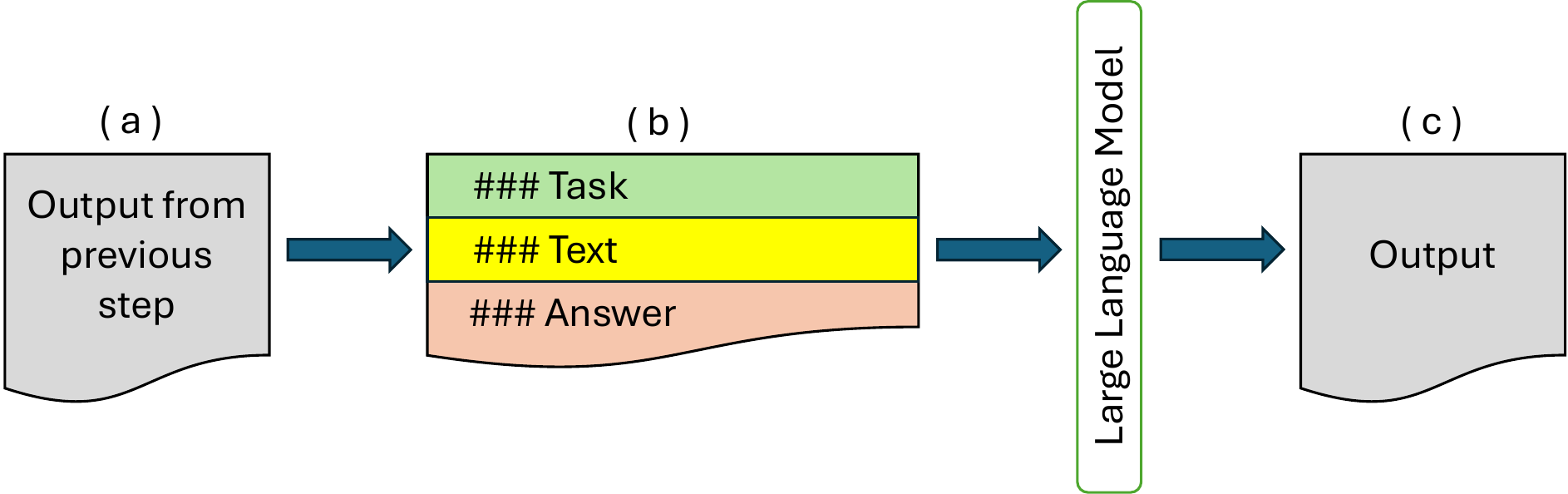}
    \caption{Pipeline for classifying the output into ``Equivalent'', ``Non-Equivalent'', and ``Unknown'' classes. Here (a) is the final output text generated by the LLMs from the prompting strategies discussed in Section 4, (b) is the overall prompt structure where $\#\#\#\: Text$ is replaced by output text described in (a), and (c) is the classification result generated by LLM.}
    \label{P:method_stage_2}
\end{figure}

In this work, rather than using the previously described approaches, we use another LLM -- from the \gpt family -- as a classifier (Figure \ref{P:method_stage_2}). We construct a prompt (see Figure~\ref{app:P:final_result}) that includes the output text generated by various prompting strategies discussed in Sections 4.1 and 4.2, along with a description of the task, to classify the text into three classes: \textit{``Equivalent''}, \textit{``Non-Equivalent''}, and \textit{``Unknown''}. The \textit{``Unknown''} class is used for cases where the LLM is unable to definitively classify the given text as either \textit{``Equivalent''} or \textit{``Non-Equivalent''}. \textcolor{black}{The output from this step is used to evaluate the effectiveness of the overall approach. Meanwhile, the output of the previous step can serve as an explanation for the final result when applying this approach in a classroom setting.}

Note that we treat the \textit{``Unknown''} class as a negative result. This approach allows us to leverage the capabilities of \gpt to streamline the classification process, reducing the need for costly and time-consuming manual evaluations.\\

\begin{figure}[!ht]
    \centering
    \begin{tcolorbox}[title=Classifying prompt]
    $\#\#\#$ Task\\
    You are given a text and you have to determine whether the given text is concluding to "Equivalent", "Non Equivalent", or "Unknown".
    Choose one of the following options ["Equivalent", "Non Equivalent", or "Unknown"]\\ \\
    $\#\#\#$ Text \\
    $<$Output from LLM$>$ \\ \\
    $\#\#\#$ Answer \\
    \end{tcolorbox}
    \caption{Prompt used to infer the final result.}
    \label{app:P:final_result}
\end{figure}

\section{Datasets}
\label{app:dataset}
\subsection{\spiderdin}
\label{app:spider_din}

\begin{table}[!tp]
    \centering
    \resizebox{0.7\textwidth}{!}{
    \begin{tabular}{cccc}
    \toprule
    \textbf{Dataset} & \textbf{Equivalent} & \textbf{Non-Equivalent} & \textbf{Exact-Match}\\
    \midrule
     \calcite & 232 & 0 & 0\\
     \spiderdin & 385 & 189 & 460\\          
     \bottomrule
    \end{tabular}
    }
    \caption{Benchmark data characteristics. Note that there are no non-equivalent pairs in \calcite. From the Spider benchmark, 460 query pairs were excluded in our \spiderdin benchmark since the generated SQL exactly matched the ground truth.}
    \label{app:tab:cal_spid_ground_truth}
\end{table}

\begin{figure*}[t]
    \centering
    \begin{subfigure}[b]{0.45\textwidth}
        \centering
        \includegraphics[width=\textwidth]{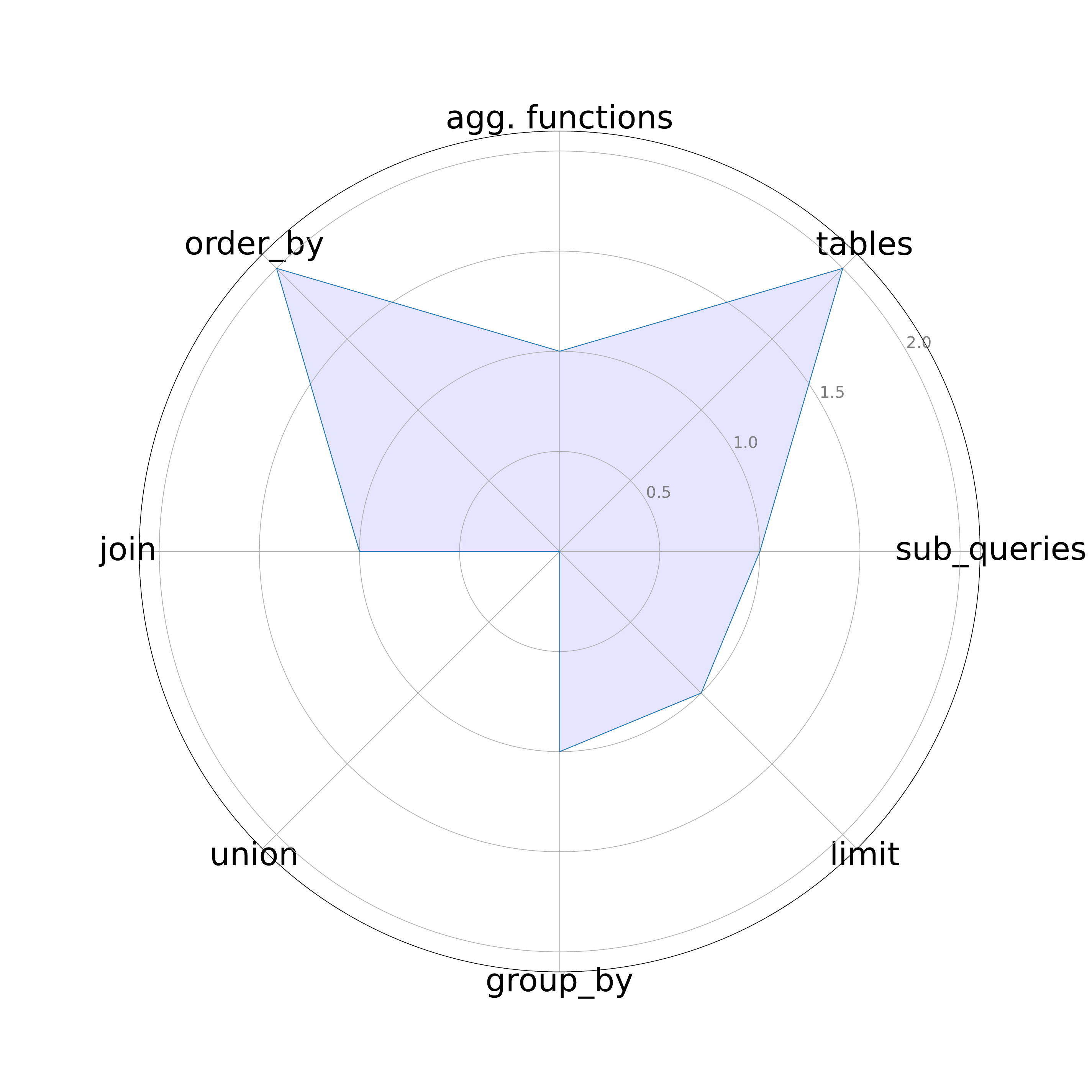}
        \caption{Question 1 and 2}
        \label{app:fig:image1_q1}
    \end{subfigure}
    \hfill
    \begin{subfigure}[b]{0.45\textwidth}
        \centering
        \includegraphics[width=\textwidth]{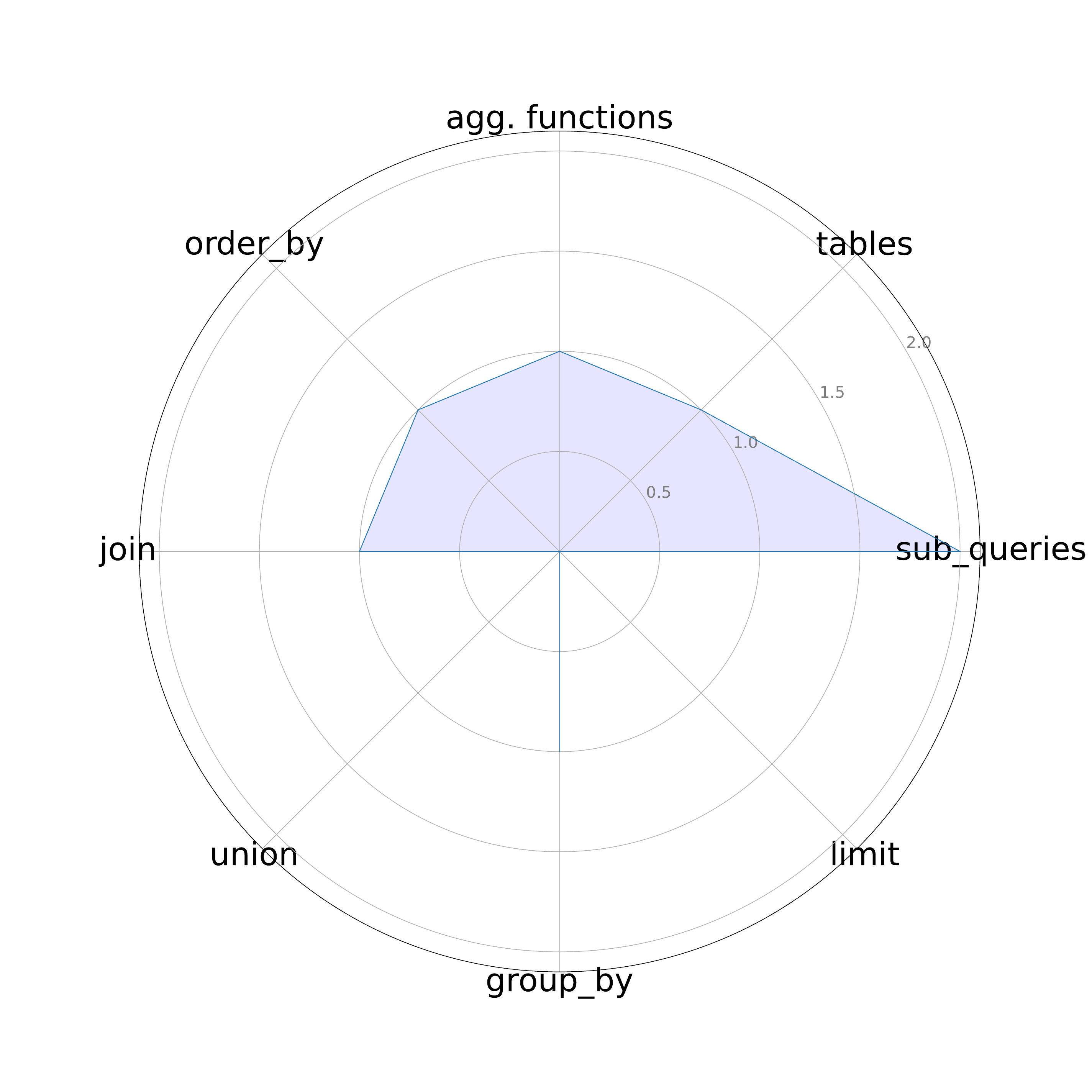}
        \caption{Question 3}
        \label{app:fig:image2_q2}
    \end{subfigure}

    \vskip\baselineskip 

    \begin{subfigure}[b]{0.45\textwidth}
        \centering
        \includegraphics[width=\textwidth]{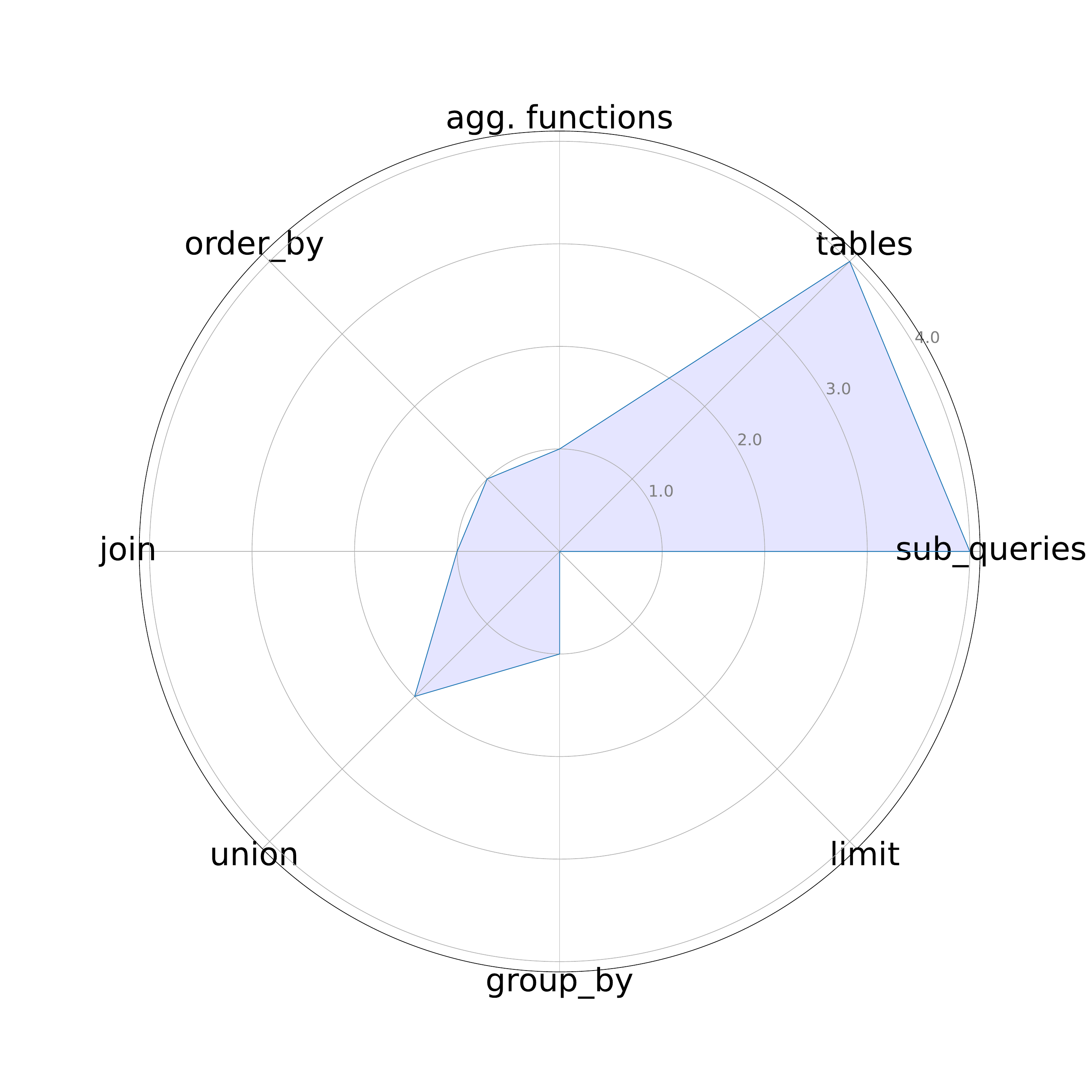}
        \caption{Question 4}
        \label{app:fig:image3_q5}
    \end{subfigure}
    \hfill
    \begin{subfigure}[b]{0.45\textwidth}
        \centering
        \includegraphics[width=\textwidth]{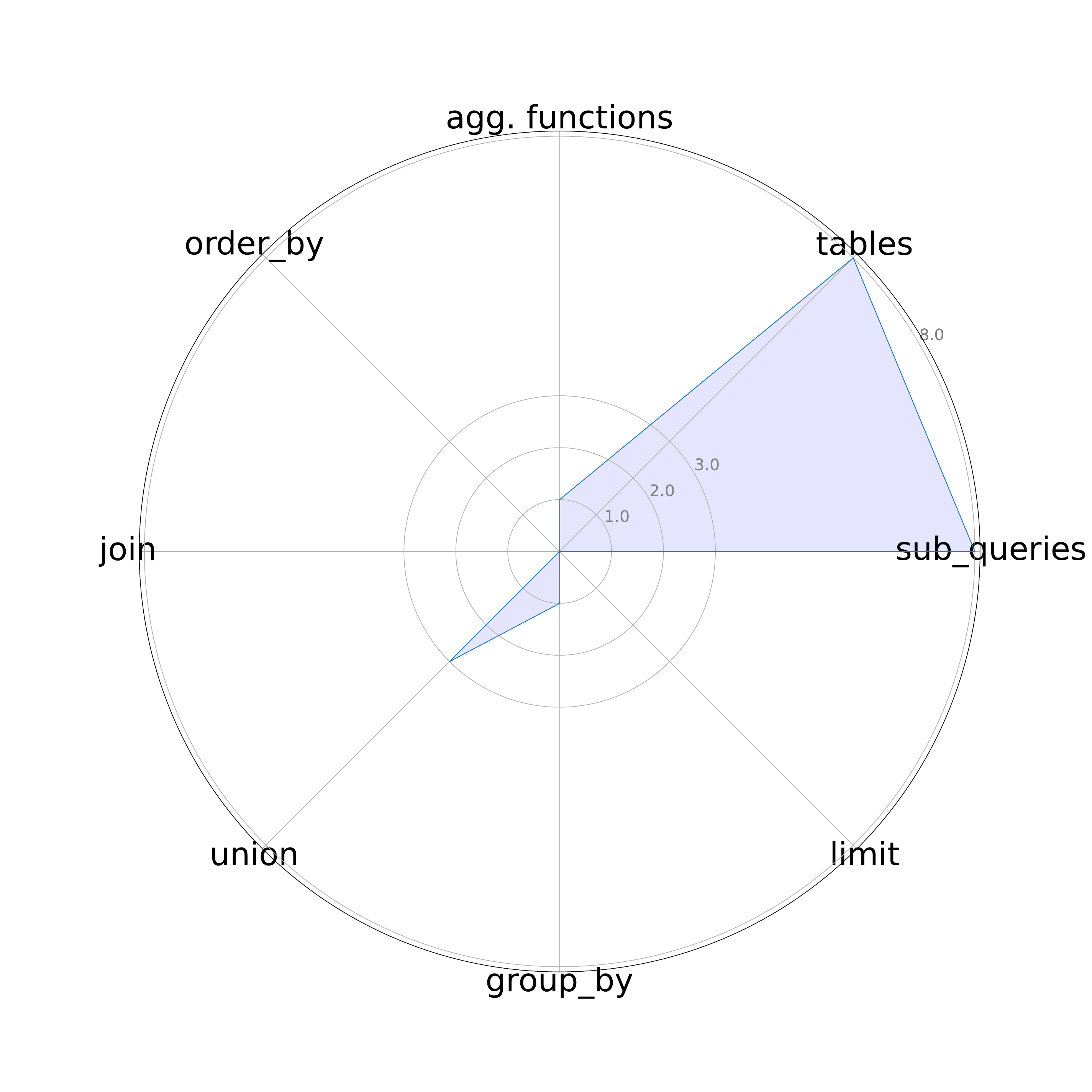}
        \caption{Question 5}
        \label{app:fig:image4_24}
    \end{subfigure}

    \caption{Distribution of various constraints in \textit{\exams} dataset. The radial axis represents the average number of constraints in the ground truth SQL query}
    \label{app:F:exams_iitd_query_complexity}
\end{figure*}

The Spider benchmark~\cite{yu2018spider} is a widely recognized benchmark primarily used for tasks related to semantic parsing and text-to-SQL generation ~\cite{Gao2023TexttoSQLEB,Pourreza2023DINSQLDI,Qi2022RASATIR,Cai2021SADGASD}. This dataset consists of natural language descriptions of information needs and corresponding SQL queries that fulfill those needs. Inspired by the approach in ~\cite{Zhao2023LLMSQLSolverCL}, our study leveraged the DIN-SQL framework \cite{Pourreza2023DINSQLDI}.
    
Using DIN-SQL, we generated an additional set of SQL queries corresponding to each sample in the Spider dataset. To evaluate the equivalence between the ground truth SQL queries and the generated SQL queries, we used test-suit-sql-eval \cite{test-suite-sql-eval}, which determines equivalence based on execution accuracy and serves as the benchmark for our study. We term this dataset as \spiderdin in the rest of the paper.

In the generated dataset, we have 1034 SQL query pairs ( Table \ref{app:tab:cal_spid_ground_truth} ). Among these pairs, 460 exhibit exact matches between the ground truth and generated SQL queries. The remaining 574 pairs do not match exactly. Within this subset of 574 non-matching pairs, 385 are classified as equivalent, while 189 are classified as non-equivalent.

\subsection{\exams}
\label{app:ground_truth_queries}
In this paper, we introduce a novel and large real-world dataset, which we call \exams, which includes five complex SQL questions that were part of a take-home assignment for the DBMS course taught at the university during the spring semester of 2023-24.

We selected five questions from different sections of the assignment (i.e., Questions 1, 2, 3, 4, and 5), each with increasing difficulty levels. The ground-truth queries were written by the instructors and verified by the teaching assistants. These queries together represent SQL queries with sophistication and complexity that are encountered in real-world settings but are rarely present in other benchmarks.  The distribution of various features of SQL covered in these SQL queries is summarized in Figure \ref{app:F:exams_iitd_query_complexity}.\\

\noindent
{\large \textbf{\exams ground truth queries:}}
\begin{itemize}
    \item Question 1: Show top 10 batsmen with highest total Caught stealing count in their entire lifetime.

    \begin{itemize}
        \item Output format: playerid, firstname, lastname, total caught stealing
        \item Order by: 1. total caught stealing (descending order), 2. firstname (ascending order), 3. lastname (ascending order), 4. playerid (ascending order)
    \end{itemize}
\end{itemize}

\begin{lstlisting}[style=Sqlstyle]

    -- Table: "Batting"
    -- Columns: "playerID", "yearID", "CS"
    
    -- Table: "People"
    -- Columns: "playerID", "nameFirst", "nameLast"
    
    with result as (
        select 
            playerid, 
            sum(coalesce(cs,0)) as total_caught_stealing 
        from batting 
        GROUP BY batting.playerid 
        order by total_caught_stealing desc 
    ) 
      
    select 
        result.playerid, 
        coalesce(people.namefirst,'') as firstname, 
        coalesce(people.namelast,'') as lastname, 
        result.total_caught_stealing 
    from result 
    join people on result.playerid = people.playerid 
    order by 
        total_caught_stealing desc, 
        namefirst asc, 
        namelast asc, 
        playerid asc 
    limit 10;
\end{lstlisting}

\begin{itemize}
    \item Question 2: Show details of top 10 batsmen with highest runscore calculated for their entire career (runscore for player’s entire career is the sum of runscores across all games he played).

    \begin{itemize}
        \item Output format: playerid, firstname, runscore.
        \item Order by: 1. runscore (descending order), 2. firstname (descending order), 3. playerid (ascending order)
    \end{itemize}
    
\end{itemize}

\begin{lstlisting}[style=Sqlstyle]

    -- Table: "Batting"
    -- Columns: "playerID", "h2b", "h3b", "hr"
    
    -- Table: "People"
    -- Columns: "playerID", "nameFirst", "nameLast"
    
    with result as (
        select 
            sum(2*coalesce(h2b,0) + 3*coalesce(h3b,0) + 4*coalesce(hr,0)) as runscore, 
            playerid 
        from batting 
        GROUP BY playerid 
        ORDER BY runscore desc 
    ) 
      
    select 
        result.playerid, 
        coalesce(people.namefirst,'') as firstname, 
        result.runscore 
    from result 
    join people on result.playerid = people.playerid 
    order by 
        result.runscore desc, 
        firstname desc, 
        playerid asc 
    limit 10;
\end{lstlisting}

\begin{itemize}
    \item Question 3: For each player show the name of the player and total points received by them from the year 2000 and later.

    \begin{itemize}
        \item Output format: playerid, playername, total points.
        \item Order by: 1. total points (descending order), 2. playerid (ascending order)
    \end{itemize}
    
\end{itemize}

\begin{lstlisting}[style=Sqlstyle]

    -- Table: "AwardsShareManagers"
    -- Coulumns: "playerID", "pointsWon" "yearID"
    
    -- Table: "People"
    -- Columns: "playerID", "nameFirst", "nameLast"
    
    with result as (
        select 
            playerid,
            sum(coalesce(pointsWon,0)) as total_points 
        from awardsshareplayers 
        where yearid>=2000 group by playerid 
    ), 
    
    player_name_table as (
        select 
            playerid,
            namefirst,
            namelast,
            coalesce(namefirst, '') || CASE WHEN (namefirst || namelast) is not NULL THEN ' ' ELSE '' END || coalesce(namelast, '') as playername 
        from people 
    )
      
    select 
        result.playerid,
        PN.playername,
        result.total_points 
    from result 
    join player_name_table PN on result.playerid = PN.playerid 
    order by 
        total_points desc,
        playerid asc;
\end{lstlisting}

\begin{itemize}
    \item Question 4: Output players with the number of seasons they have played in decreasing order. Note that a player might have played multiple roles out of batter, pitcher, and fielder in the same season; it still counts as one season.

    \begin{itemize}
        \item Output format: playerid, firstname, lastname, date of birth, num seasons.
        \item Order by: 1. num seasons (descending order) 2. playerid (ascending order) 3. firstname (ascending order) 4. lastname (ascending order) 5. date of birth (ascending order)
    \end{itemize}
    
\end{itemize}

\begin{lstlisting}[style=Sqlstyle]

    -- Table: "Batting"
    -- Columns: "playerID", "yearID", "CS"
    
    -- Table: "Fielding"
    -- Columns: "playerID", "yearID"
    
    -- Table: "Pitching"
    -- Columns: "playerID", "yearID"
    
    -- Table: "People"
    -- Columns: "playerID", "nameFirst", "nameLast"
    
    with all_tables as (
        select 
            playerid,
            yearid 
        from batting 
        group by 
            playerid, 
            yearid 
        union 
        select 
            playerid, 
            yearid 
        from fielding 
        group by 
            playerid, 
            yearid 
        union 
        select 
            playerid, 
            yearid 
        from pitching 
        group by 
            playerid, 
            yearid
    ), 
    
    result as (
        select 
            playerid, 
            count(distinct(yearid)) as num_seasons 
        from all_tables 
        group by 
            playerid 
    ) 
      
    select 
        r.playerid,
        coalesce(p.namefirst,'') as firstname,
        coalesce(p.namelast,'') as lastname,
        coalesce(birthyear || '-' || lpad(birthmonth::text,2,'0') || '-' || lpad(birthday::text,2,'0'),'') as date_of_birth,
        r.num_seasons 
    from result r 
    join people p on r.playerid = p.playerid 
    order by 
        r.num_seasons desc,
        r.playerid asc;
\end{lstlisting}

\begin{itemize}
    \item Question 5: Graph 1: The concept of graphical analysis can be applied to this dataset. Consider the tables pitching and allstarfull. Now, we can make a graph such that all the players who appear in these tables are nodes of this graph. The edges are defined such that there exists an edge between two nodes (players) if they have played in the same team in the same season; the weight of the edge is the number of seasons played together in the same team. The graph thus formed is undirected and weighted. If players A and B played 5 seasons together in team X and 2 seasons together in team Y, the graph will have one edge between A and B with weight 7. For allstarfull table, consider only the tuples where the player actually played in the game i.e., GP = 1.\\

    Query: Using Graph 1, find whether there exists a path of length three or more between webbbr01 and clemero02. Output a boolean value: True for yes and False for no.

    \begin{itemize}
        \item Output format: pathexists
    \end{itemize}

\end{itemize}

\begin{lstlisting}[style=Sqlstyle]
    -- Table: "AllStarFull"
    -- Columns: "playerID", "YearID", "teamID", "GP"
    
    -- Table: "Pitching"
    -- Columns: "playerID", "yearID", "teamID"
    
    create or replace view graph1 as 
        select 
            p1, 
            p2, 
            count(*) as w 
        from ( 
            select table1.playerid as p1, 
            table2.playerid as p2, 
            table1.teamid, 
            table1.yearid 
            from ( 
                select 
                    * 
                from allstarfull 
                where GP = 1 
            ) as table1, 
            ( 
                select 
                    * 
                from allstarfull 
                where GP = 1
            ) as table2 
        where 
            table1.teamid = table2.teamid and 
            table1.yearid = table2.yearid and 
            not table1.playerid = table2.playerid 
        union 
        select 
            table1.playerid as p1, 
            table2.playerid as p2, 
            table1.teamid, 
            table1.yearid 
        from 
            pitching as table1, 
            pitching as table2 
        where 
            table1.teamid = table2.teamid and 
            table1.yearid = table2.yearid and 
            not table1.playerid = table2.playerid 
        union 
        select 
            table1.playerid as p1, 
            table2.playerid as p2, 
            table1.teamid, 
            table1.yearid 
        from 
            pitching as table1, 
            (
                select 
                    * 
                from allstarfull 
                where GP = 1
            ) as table2 
        where 
            table1.teamid = table2.teamid and 
            table1.yearid = table2.yearid and 
            not table1.playerid = table2.playerid 
        union 
        select 
            table1.playerid as p1, 
            table2.playerid as p2, 
            table1.teamid, table1.yearid 
        from (
                select 
                    * 
                from allstarfull 
                where GP = 1
            ) as table1, 
            pitching as table2 
        where 
            table1.teamid = table2.teamid and 
            table1.yearid = table2.yearid and 
            not table1.playerid = table2.playerid 
        ) as temp 
        group by 
            p1, 
            p2; 
    
    with recursive sub as (
        select 
            array[p1::text, p2::text] as path, 
            p2, 
            w as dist 
        from graph1 
        where p1 = 'webbbr01' 
        union all 
        select 
            recur.path || graph1.p2::text, 
            graph1.p2, 
            (dist + w) as dist 
        from 
            sub as recur, 
            graph1 
        where 
            graph1.p1 = recur.p2 and 
            not graph1.p2 = any (recur.path) and 
            not recur.p2 = 'clemero02' 
    ) 
      
    select 
        case when count(*) > 0 then True else False end as pathexists 
    from sub 
    where 
        dist >= 3 and 
        p2 = 'clemero02';
\end{lstlisting}

\section{Experimental Setup \& Hyper-parameters}
\subsection{Inference using LLMs}
All experiments were conducted using the Ubuntu 18.04 LTS operating system on an NVIDIA DGX station equipped with four V100-SXM2 GPU cards, each having 32GB of GPU memory. The system also included 256GB of RAM and a 64-core Intel(R) Xeon(R) Gold 5218 CPU @ 2.30GHz. For chosen LLMs, the details of the hyper-parameter choices used in our experiments are as follows:

\begin{itemize}
    \item \textbf{\codellama}: We used two variants of the \codellama model --  \codellama-7B-Instruct-hf and \codellama-13B-Instruct-hf, which have 7 billion and 13 billion parameters, respectively. The hyper-parameters of these models were configured as follows: \textit{`max\_new\_tokens'} was set to 500, \textit{`num\_beams'} to 1, and \textit{`do\_sample'} to False. Results presented in Table \ref{T:all} and Table \ref{T:exams_iitd} are derived without using any quantization techniques.
    \item \textbf{\gpt}: We used two variants of \gpt \space models -- \gpt-3.5-turbo-0125 and \gpt-4-0125-preview. These models were configured with a \textit{`temperature'} of 0.2~\footnote{lower values of \textit{temperature} parameter resulting in more consistent outputs from \gpt.}, \textit{`max\_tokens'} of 1000, and \textit{`frequency\_penalty'} of 0.0.
    \item \textbf{\geminipro}: For \geminipro we used a \textit{`temperature'} setting of 0.2, \textit{`max\_output\_tokens'} capped at 10000, and \textit{`top\_k value'} of 1.
\end{itemize}

\subsection{Fine-tuning of LLMs}
\label{app:finetuneing}

\raj{The experiments presented in Sections \ref{performance_of_llm} and \ref{performance_of_pre_trained_llms_against_query_complexity} use standard LLMs without task-specific fine-tuning. In this section, we present the impact of fine-tuning LLMs for task-specific objectives. Specifically, we aim to evaluate how adapting LLMs to the nuances of SQL query equivalence affects their performance and accuracy in this specialized domain. Due to resource limitations, the fine-tuning process was performed using 8-bit quantization.}



\begin{table}[!t]
    \centering
    \resizebox{1.0\textwidth}{!}{
        \begin{tabular}{cl cc cc cc cc cc c}
        \toprule
        \multirow{2}{*}{\textbf{Prompting}} & \multirow{2}{*}{\textbf{Models}} & \multicolumn{2}{c}{\textbf{Easy}} & \multicolumn{2}{c}{\textbf{Medium}} & \multicolumn{2}{c}{\textbf{Hard}} & \multicolumn{2}{c}{\textbf{Extra Hard}} & \multicolumn{2}{c}{\textbf{Total}} & \multirow{3}{*}{\textbf{GM}}\\
        \cmidrule(lr){3-4}\cmidrule(lr){5-6}\cmidrule(lr){7-8}\cmidrule(lr){9-10}\cmidrule(lr){11-12} & & EQ & NEQ & EQ & NEQ & EQ & NEQ & EQ & NEQ & EQ & NEQ\\
        & \multicolumn{1}{c}{\textit{Ground truth}} & 38 & 20 & 188 & 62 & 84 & 42 & 75 & 65 & 385 & 189\\
        
        \midrule
        
        Basic & \codellama-13B & 38 & 0 & 188 & 0 & 84 & 0 & 74 & 1 & 0.997 & 0.005 & 0.0707 \\
        Prompting & \codellama-13B 8bit finetune & 34 & 10 & 163 & 15 & 79 & 8 & 69 & 12 & 0.896 & 0.238 & 0.4618 \\
        
        \cmidrule(lr){1-13}

        Basic Prompting & \codellama-13B & 38 & 0 & 187 & 0 & 84 & 1 & 74 & 1 & 0.995 & 0.011 & 0.1046 \\
        with LP & \codellama-13B 8bit finetune & 26 & 5 & 133 & 26 & 61 & 12 & 35 & 27 & 0.662 & 0.370 & \textbf{0.4949} \\
        \bottomrule
        \end{tabular}
    }
    \caption{Performance on \textit{spider+DIN} using basic prompting \& basic prompting with LP; with and without fine-tuning. Here, EQ refers to \textit{Equivalent}, NEQ refers to \textit{Non-Equivalent}, GM refers to \textit{Geometric Mean}, and LP refers to \textit{Logical Plan} .}
    \label{T:finetune}
\end{table}

\textbf{Dataset}: For dataset preparation, we used the dataset curated by ~\cite{He2024VeriEQLBE}, which comprises 23,994 query pairs. This dataset was curated by collecting all publicly available queries accepted by LeetCode\footnote{LeetCode. 2023. LeetCode website. https://leetcode.com/} through web crawling. Among these queries, 62.1\% (14,905 query pairs) were labeled as equivalent, 14.9\% (3,586 query pairs) were labeled as non-equivalent, and the remaining 23\% (5,503 query pairs) were labeled as unsupported. For this comprehensive dataset, we randomly sampled 1,000 equivalent and 1,000 non-equivalent query pairs to fine-tune the model.  This sampling strategy ensures a balanced representation of equivalent and non-equivalent queries, providing a robust foundation for model training and evaluation. Further, we sampled 400 equivalent and 400 non-equivalent query pairs as the evaluation set.

\noindent
\textbf{Fine-tuning}: We fine-tuned the \codellama-13B model for the task of SQL query equivalence checking using parameter-efficient fine-tuning (PEFT) with low-rank adaptation (LoRA). LoRA operates by fixing the original weights of the model and introducing trainable low-rank matrices into each layer of the transformer architecture. This approach significantly reduces the number of trainable parameters, thereby decreasing the computational load while enhancing the model's performance on downstream tasks. The parameters for fine-tuning are set as follows:
    
\begin{center}
    \begin{tabular}{|p{0.45\columnwidth}p{0.45\columnwidth}|}
    \hline
     & \\
    r: 16 & warmup\_steps: 100 \\
    lora\_alpha: 16 & max\_step: 400\\
    lora\_dropout: 0.05 & learning\_rate: 3$e^{-4}$ \\
    optim: adamw\_torch & \\ & \\
    \hline
    \end{tabular}
    
\end{center}

\noindent
\textbf{Resource}: For fine-tuning, we used the Ubuntu 20.04 LTS operating system on an NVIDIA DGX station equipped with four A100-PCIE GPUs, each having 40GB of GPU memory. The system also includes 515GB of RAM and a 96-core  Intel(R) Xeon(R) Gold 6248R CPU @ 3.00GHz. For fine-tuning, we used two A100-PCIE GPUs, which took around 20 hours (~1200 minutes) to reach convergence.

\noindent
\textbf{Input}: Since the dataset curated by \cite{Gao2023TexttoSQLEB} does not provide any explanations for why a given SQL query pair is classified as equivalent or non-equivalent. Consequently, we used the basic prompting method to generate the input prompts for fine-tuning the models. Additionally, we evaluated the performance of these fine-tuned models on the \textit{spider+DIN} dataset using the same basic prompting approach with and without a query logical plan. 

\noindent
\textbf{Result:} Table \ref{T:finetune} presents the performance of the fine-tuned \codellama-13B model on the \textit{spider+DIN} dataset. Fine-tuning significantly improved the model's accuracy in classifying \textit{non-equivalent} SQL query pairs, while performance on \textit{equivalent} pairs remained consistent with a minor decrease. Overall, the fine-tuned model demonstrated a notable performance improvement.


\end{document}